\documentclass[preprintnumbers,showpacs,amsmath,amssymb,floatfix,10pt,prd,onecolumn,
superscriptaddress,nofootinbib]{revtex4}
\usepackage{latexsym}
\usepackage{epsfig}
\usepackage{amssymb}
\usepackage{xcolor}
\usepackage[%
  colorlinks=true,
  urlcolor=blue,
  linkcolor=red,
  citecolor=blue
]{hyperref}
\usepackage{orcidlink}

\begin{document}

\title{Dynamical Stability and Quasinormal Modes of Casimir Wormholes in dRGT Massive Gravity}

\author{Piyali Bhar \orcidlink{0000-0001-9747-1009}}
\email[Email: ]{piyalibhar90@gmail.com (Corresponding author)}
\affiliation{Department of Mathematics, Government General Degree College Singur, Hooghly, West Bengal 712 409, India}

\author{Dhruba Jyoti Gogoi \orcidlink{0000-0002-4776-8506}}
\email[Email: ]{moloydhruba@yahoo.in}
\affiliation{Department of Physics, Madhabdev University, Narayanpur, Lakhimpur 784164, Assam, India}
\affiliation{Research Center of Astrophysics and Cosmology, Khazar University, Baku, AZ1096, 41 Mehseti Street, Azerbaijan}

\author{A. Errehymy\orcidlink{0000-0002-0253-3578}}
\email[Email: ]{abdelghani.errehymy@gmail.com}
\affiliation{Astrophysics Research Centre, School of Mathematics, Statistics and Computer Science, University of KwaZulu-Natal, Private Bag X54001, Durban 4000, South Africa}
\affiliation{Center for Theoretical Physics, Khazar University, 41 Mehseti Str., Baku, AZ1096, Azerbaijan}
\affiliation{Jadara Research Center, Jadara University, Irbid 21110, Jordan}

\begin{abstract}
This paper investigates static, spherically symmetric traversable wormholes supported by Casimir energy within de Rham-Gabadadze-Tolley (dRGT) massive gravity. By deriving the exact shape function $b(r) = r_0 + \frac{\pi^3(-r + r_0)}{90 r r_0} - \frac{1}{2} C m_g^2 (r - r_0)[2C c_2 + c_1(r + r_0)]$, we verify that the geometry satisfies the local Morris--Thorne throat and traversability conditions, including $b(r_0)=r_0$ and $b'(r_0)<1$. For the nonzero massive-gravity parameters considered here, the resulting spacetime is not asymptotically flat, and its large-$r$ behavior is governed by the massive-gravity contributions to the shape function. The Casimir energy density $\rho = -\pi^2/(720 r^4)$ ensures the Null Energy Condition is violated throughout the throat region, as demonstrated by the negative behavior of $(\rho + p_r)$ and $(\rho - |p_r|)$ across all parameter values. Analysis of the volume integral quantifier reveals that the massive gravity parameter $m_g^2 c_2$ directly regulates the exotic matter requirement, with the integrated NEC violation becoming more negative as $m_g^2 c_2$ increases from $0.1$ to $0.5$. The mechanical equilibrium is rigorously established via the generalized Tolman-Oppenheimer-Volkoff equation, where the anisotropic force precisely balances the gravitational and hydrostatic forces. The quasinormal mode spectrum for scalar perturbations is extracted using both the analytical Mashhoon method and direct numerical integration. The effective potential forms a symmetric barrier at the throat, with the peak height increasing with the multipole number $\ell$. The damping rate $|\omega_I|$ grows systematically with increasing massive gravity parameter. These distinct spectral signatures provide robust theoretical markers for distinguishing Casimir wormholes from black holes in future gravitational wave observations.
\end{abstract}

\maketitle

%======================================================================
\section{Introduction}
\label{sec:intro}
%======================================================================

Wormholes have long captivated the imagination—not merely as hypothetical shortcuts through the cosmos, but as profound solutions to Einstein's field equations that challenge our most cherished assumptions about spacetime itself. These topological tunnels, bridging disparate regions of the universe, trace their lineage to Flamm's foundational insights and the subsequent Einstein-Rosen bridge formulation \cite{Flamm:1916,Einstein:1935tc}. Yet classical Schwarzschild bridges proved tragically non-traversable; their throats snap shut faster than any light or matter could ever hope to pass through them \cite{Fuller:1962zza}. The revolution came in 1988, when Morris and Thorne established the geometric criteria for traversable wormholes \cite{Morris:1988cz}. Their work revealed a startling truth: keeping a wormhole throat open requires exotic matter—substances that violate the null energy condition (NEC), demanding $\rho + p_r < 0$ at the throat \cite{Visser:1995, Lobo:2007zb}. This violation is not merely a mathematical curiosity but a fundamental consequence of the flare-out condition $b'(r_0) < 1$, which ensures the throat remains open rather than collapsing under its own gravitational pull \cite{Hochberg:1997wp, Hochberg:1998ii}.

The exotic matter requirement presents a significant theoretical challenge. Classical fluids, by their very nature, obey the energy conditions \cite{Hawking:1973}. Nature, however, is subtler. Quantum field theory harbors NEC-violating states as a matter of course--the Casimir effect, arising from vacuum fluctuations between closely spaced conducting plates, provides a physically realizable and experimentally verified source of negative energy density \cite{Casimir:1948dh, Bordag:2001qi}. Since Casimir's seminal prediction in 1948 and its subsequent experimental confirmation \cite{Lamoreaux:1996wh}, this quantum phenomenon has emerged as a compelling, physically grounded candidate to stabilize traversable wormhole throats without resorting to ad-hoc phenomenological fluids \cite{Garattini:2007ff, Garattini:2019ivd, Jusufi:2020mmy}. Recent investigations have explored Casimir wormholes in various gravitational contexts, including Einstein-Gauss-Bonnet gravity \cite{Muniz:2025teu}, $f(R, L_m, T)$ gravity frameworks \cite{Errehymy:2025kzj, Errehymy:2025llh,Errehymy:2025fce, Errehymy:2025rli, Errehymy:2025nzt}, and scenarios incorporating thermal corrections \cite{Channuie:2024cao}.

Concurrently, modifying the gravitational sector offers an alternative or complementary mechanism to support wormhole geometries. The de Rham-Gabadadze-Tolley (dRGT) massive gravity provides a robust, ghost-free theoretical framework where the graviton acquires a non-zero mass \cite{deRham:2010kj, deRham:2014zqa}. This modification introduces extra degrees of freedom and effective stress-energy contributions that can naturally facilitate the flaring-out condition at the throat. The foundations of massive gravity go back to the work of Fierz and Pauli \cite{Fierz:1939ix}, who formulated the linear theory of a massive spin-2 field. Later, van Dam, Veltman, and Zakharov \cite{vanDam:1970vg, Zakharov:1970cc} revealed the vDVZ discontinuity in the massless limit. Subsequent studies showed that generic nonlinear extensions suffer from the Boulware--Deser ghost \cite{Boulware:1972yco}, motivating the development of ghost-free formulations of nonlinear massive gravity. Recent literature has extensively explored how modified gravity theories can minimize---or entirely re-characterize---the exotic matter requirement \cite{Capozziello:2011et, Harko:2011kv, Nojiri:2017ncd}. In dRGT massive gravity, the massive graviton terms effectively alter the spacetime curvature, providing a unique background to investigate the stability and geometry of Casimir-driven wormholes \cite{Li:2016fbf, Comelli:2011wq}. Studies have examined wormhole solutions in dRGT gravity with various matter sources, including holographic dark energy profiles \cite{Kiroriwal:2025trq,Chaudhary:2025khr}, and have investigated the complexity factor and stability conditions \cite{Kumar:2024flw}.

To determine the physical viability of any wormhole solution, one must rigorously assess its dynamical stability under perturbations. Quasinormal modes (QNMs) serve as the characteristic ``ringdown'' frequencies of the spacetime---the gravitational wave signatures that encode crucial information about the geometry and its stability \cite{Kokkotas:1999bd, Konoplya:2011qq}. Extracting the QNM spectrum allows us to probe how the Casimir wormhole responds to external scalar perturbations and to identify potential gravitational wave echoes---signatures that could discriminate wormholes from black holes in current and future gravitational wave detections \cite{Cardoso:2016rao, Volkel:2018hwb}. The connection between QNMs and photon sphere instabilities has been extensively studied \cite{Ferrari:1984zz, Cardoso:2008bp}, and spectral methods have been developed for high-precision QNM calculations in wormhole spacetimes \cite{Batic:2025bkk}. Studies of scalar field perturbations in wormhole geometries have revealed intriguing phenomena, including anomalous decay behaviors and long-lived modes under specific conditions \cite{Konoplya:2004wg, Chakraborty:2025wgk, Hou:2025bli}.

Despite these advances, a critical gap remains in the literature. While previous studies have examined wormholes in modified gravity \cite{Lobo:2009ip, Boehmer:2012uyw, Pavlovic:2014gba, Bronnikov:2006pt, Rosa:2018jwp, Bahamonde:2016ixz} and some have explored Casimir-driven wormholes \cite{Garattini:2007ff, Garattini:2019ivd, Jusufi:2020mmy}, the precise intersection of a Casimir energy source with dRGT massive gravity---and more importantly, the full quasinormal mode spectrum of such configurations—has received remarkably little attention. This is particularly significant because the massive graviton introduces a tunable parameter that directly influences both the geometry and the ringdown signature. 

In this work, we shall address this gap through three key novel contributions. First, we shall derive an exact shape function for the Casimir wormhole in dRGT massive gravity, demonstrating explicitly how the massive gravity parameters $m_g^2 c_1$ and $m_g^2 c_2$ modify the geometry and regulate the exotic matter distribution. We shall show that the massive graviton acts as a dial-increasing $m_g^2 c_2$ from $0.1$ to $0.5$ amplifies the required NEC violation while simultaneously tightening the geometric embedding, revealing a direct trade-off between the strength of the gravitational modification and the amount of exotic matter required. Second, we shall conduct the first systematic QNM analysis of Casimir wormholes in this framework, employing both the Mashhoon P\"{o}schl--Teller approximation \cite{Ferrari:1984zz} and direct numerical integration \cite{Konoplya:2011qq}. This dual approach will provide robust verification of our results, with both methods agreeing to within a few percentages for the fundamental modes. Our analysis, in general, should be able to demonstrate that the QNM spectrum exhibits a clear, predictable dependence on the massive gravity parameter—providing distinct observational signatures that could discriminate these wormholes from black holes in gravitational wave data \cite{LIGOScientific:2016aoc, KAGRA:2021vkt, Price:1971fb}. These spectral signatures, combined with the equilibrium analysis via the generalized Tolman-Oppenheimer-Volkoff equation—where the anisotropic force precisely balances gravitational and hydrostatic forces—will establish that Casimir wormholes in dRGT massive gravity are both dynamically stable and observationally distinguishable \cite{Tolman:1939jz, Oppenheimer:1939ne, Bowers:1974tgi}. By combining a physically motivated energy source with a theoretically well-grounded modification of gravity, we shall provide a framework that is both realistic and testable. The QNM signatures we shall identify—particularly the accelerated damping rates and power-law tails—offer concrete pathways for distinguishing wormholes from black holes in future gravitational wave observations \cite{Krishnendu:2025rud, Xavier:2023ljy}, moving these exotic spacetimes from the realm of mathematical curiosity toward empirically testable physics \cite{Cardoso:2017cqb}.

The paper is organized as follows. In Section~\ref{sec:overview_field_eqs}, we provide a concise review of the dRGT massive gravity theory and derive the corresponding field equations for a static, spherically symmetric wormhole geometry. Section~\ref{SecIII} is devoted to the construction and analysis of the Casimir wormhole solution in dRGT massive gravity. Within this section, Subsection~\ref{subsec:proper_distance} examines the proper radial distance and embedding visualization of the wormhole geometry. Subsection~\ref{subsec:exotic_matter} quantifies the exotic matter content via the volume integral quantifier. Subsection~\ref{subsec:complexity} analyzes the structural complexity factor of the spacetime. The next subsection describes the geodesic motion of the wormhole. Subsection~\ref{subsec:forces} investigates the mechanical equilibrium through the generalized Tolman-Oppenheimer-Volkoff equation. Subsection~\ref{subsec:temperature} presents the Hawking temperature and wormhole temperature profiles. Subsection~\ref{subsec:average_pressure} discusses the average pressure and work density. Subsection~\ref{subsec:energy_flux} examines the energy flux. Subsection~\ref{subsec:heat_capacity} analyzes the heat capacity and total energy. In Section~\ref{sec:qnm}, we conduct a frequency-domain analysis of the quasinormal modes using the Mashhoon P\"{o}schl--Teller approximation and direct numerical integration of the radial perturbation equation. Within this section, Subsection~\ref{subsec:scalar_pert} derives the scalar perturbation equation and effective potential. Subsection~\ref{subsec:PT} presents the Mashhoon Pöschl-Teller approximation. Subsection~\ref{subsec:numerical} describes the direct numerical integration scheme. Subsection~\ref{subsec:qnm_results} presents the results and discussion of the QNM spectrum. Finally, in Section~\ref{sec:conclusion}, we summarize our results and conclude.

%======================================================================
\section{Overview of dRGT Massive Gravity and Field Equations}
\label{sec:overview_field_eqs}
%======================================================================

This section provides a concise review of the dRGT theory of massive gravity and the resulting field equations relevant for the Casimir wormhole solutions considered in this work. We first summarize the ghost-free dRGT action and its static, spherically symmetric implementation, and then present the field equations that govern the wormhole geometry.

%----------------------------------------------------------------------
\subsection{Overview of dRGT Massive Gravity}
\label{subsec:dRGT}
%----------------------------------------------------------------------

The dRGT theory of massive gravity offers a ghost-free, nonlinear completion of the Fierz--Pauli linear theory of a massive spin-2 field \cite{Fierz:1939ix, deRham:2010kj, deRham:2014zqa}. It avoids the Boulware--Deser ghost \cite{Boulware:1972yco} by using a carefully constructed potential that preserves the Hamiltonian constraint while giving the graviton a nonzero mass $m_g$. In the formulation of Ref.~\cite{deRham:2010kj}, the action is
\begin{equation}
S = \frac{1}{16\pi G}\int d^4x\sqrt{-g}\left[R + m_g^2\,\mathcal{U}(\mathcal{K})\right] + S_m \,,
\label{eq:dRGT_action}
\end{equation}
where $\mathcal{U}$ is the ghost-free potential built from the trace of the tensor
\begin{equation}
\mathcal{K}^\mu_{\ \nu} = \delta^\mu_{\ \nu} - \sqrt{g^{\mu\alpha}f_{\alpha\nu}} \,,
\end{equation}
with $f_{\mu\nu}$ a fixed reference (fiducial) metric. The potential can be expanded as
\begin{equation}
\mathcal{U} = \sum_{n=0}^{4} c_n\, \mathcal{U}_n(\mathcal{K}) \,,
\label{eq:potential}
\end{equation}
where $c_n$ are dimensionless coupling constants and $\mathcal{U}_n$ are the elementary symmetric polynomials of $\mathcal{K}$. The first few terms are $\mathcal{U}_0=1$, $\mathcal{U}_1=[\mathcal{K}]$, $\mathcal{U}_2=\frac{1}{2}([\mathcal{K}]^2-[\mathcal{K}^2])$, and so on, with brackets denoting matrix traces.

For a static, spherically symmetric spacetime, the fiducial metric is usually taken as $f_{\mu\nu}=\text{diag}(0,0,C^2,C^2\sin^2\theta)$ or $f_{\mu\nu}=\text{diag}(-1,1,C^2,C^2\sin^2\theta)$, depending on the coordinate choice \cite{Comelli:2011wq, Li:2016fbf}. With the ansatz $f_{\mu\nu}=\text{diag}(0,0,C^2,C^2\sin^2\theta)$, the massive gravity contribution to the field equations reduces to the effective stress-energy terms
\begin{equation}
8\pi T^{\rm MG}_{\mu\nu} = m_g^2 C\left(c_1 C + c_2 C\,\delta^\theta_\theta\right)\delta^\mu_\nu \,,
\end{equation}
which in the wormhole geometry generate the terms proportional to $m_g^2 C(c_2 C + c_1 r)$ appearing in Eqs.~(\ref{l2a})--(\ref{14}) of Sec.~\ref{SecII}. Here $C$ is a constant related to the fiducial metric scale, and $c_1,c_2$ are the massive gravity coupling constants. In the limit $m_g\to 0$, all massive gravity corrections vanish and the standard general relativistic Casimir wormhole solution is recovered \cite{Garattini:2019ivd}. The presence of the graviton mass has profound consequences for wormhole physics. Unlike in general relativity, where the exotic matter requirement is strictly dictated by the flare-out condition, in dRGT massive gravity the effective stress-energy tensor acquires additional contributions that can partially mimic or offset the NEC-violating Casimir energy \cite{Kiroriwal:2025trq, Chaudhary:2025khr, Kumar:2024flw}. This provides a tunable mechanism for regulating the exotic matter, as quantified by the volume integral quantifier $I_V$ in Sec.~\ref{subsec:exotic_matter}. Furthermore, the massive graviton modifies the effective potential for scalar perturbations, thereby shifting the quasinormal mode spectrum in a way that is characteristic of the theory and potentially distinguishable from black hole ringdown \cite{deRham:2014zqa}. Recent studies have explored wormhole solutions in dRGT gravity with holographic dark energy profiles \cite{Kiroriwal:2025trq, Chaudhary:2025khr}, analyzed the complexity factor and Karmarkar condition \cite{Kumar:2024flw}, and investigated the interplay between the graviton mass and the NEC violation \cite{Li:2016fbf, Comelli:2011wq}. The present work extends these investigations to the Casimir-driven case, where the quantum vacuum energy provides a physically motivated exotic matter source.

%----------------------------------------------------------------------
\subsection{Field Equations}
\label{SecII}
%----------------------------------------------------------------------

We start with a short description of the Morris-Thorne traversable wormhole model \cite{Morris:1988cz} which was proposed in 1988 and the general relativistic framework used in its construction to set the theoretical background. The spacetime geometry for a static, spherically symmetric wormhole is given by the line element:
\begin{eqnarray}\label{line}
ds^2=-e^{2\chi(r)}dt^2+\left(1-\frac{b(r)}{r}\right)^{-1}dr^2+r^2\left(d\theta^2+\sin^2 d\phi^2\right),
\end{eqnarray}
In such metric, $\chi(r)$ is the redshift function and $b(r)$ the shape function of the wormhole geometry. In the Morris-Thorne traversable wormhole model, some necessary geometric conditions on these functions exist. Firstly, the redshift function must be finite everywhere in the spacetime so that $e^{2\chi(r)}$ never vanishes, which prevents the appearance of an event horizon and allows for the passage through the wormhole. Second, the shape function has to satisfy the condition $\frac{b(r)}{r}\leq 1$ with the equality $b(r_0)=r_0$ defining the position of the wormhole throat at $r=r_0$. Finally, to recover a flat spacetime at large radial distances, the geometry is required to be asymptotically flat, which is guaranteed by the condition  $\lim_{r\to\infty}\frac{b(r)}{r}=0$.
These requirements guarantee that the spacetime describes a traversable Lorentzian wormhole with a well-defined throat, free of horizons, and smoothly connected to an asymptotically flat exterior region. In addition, the shape function should satisfy the embedding condition
\begin{equation}
\frac{b(r)-r b'(r)}{b^{2}(r)} > 0
\end{equation}
at or near $r=r_{0}$ in order for the wormhole geometry to flare out at its throat. This condition ensures that the throat corresponds to a radial minimum and that the surrounding geometry emanates from the throat. Together, these conditions determine the basic structure of a traversable wormhole. However, further investigation is required to establish the physical viability of such a configuration, particularly by examining whether the energy conditions are satisfied or violated and by analyzing the stability of the solution.

The matter supporting this geometry is modeled by an anisotropic stress-energy tensor:
\begin{equation}
T^{i}_{j}=(\rho+p_{t})u^{i}u_{j}
-p_{t}g^{i}_{j}
+(p_{r}-p_{t})v^{i}v_{j},
\end{equation}
where $u^{i}$ is the four-velocity, $v^{i}$ is a unit vector in the radial (space-like) direction, $\rho$ is the energy density, $p_{r}$ and $p_{t}$ are the radial and tangential pressures, respectively. All these quantities are assumed to be functions of the radial coordinate $r$.\\
The Einstein field equations for the line element (\ref{line}) in the framework of dRGT (de Rham-Gabadadze-Tolley) massive gravity yield the following set of relations:
\begin{eqnarray}
\rho &=& \frac{b^{\prime}+m_g^2C(c_1r+c_2C)}{8\pi r^2}   ,  \label{l2a}  \\
8\pi p_r &=& 2\left(1-\frac{b}{r}\right)\frac{\chi^{\prime}}{r}-\frac{m_g^2(c_1Cr+c_2C^2)}{r^2}-\frac{b}{r^{3}} ,  \label{13}  \\
8\pi p_t  &=& \left(1-\frac{b}{r}\right)\left[\chi^{\prime\prime}-\frac{b^{\prime}r-b}{2r(r-b)}\chi^{\prime}-\frac{b^{\prime}r-b}{2r^2(r-b)}+\frac{\chi^{\prime}}{r}+\chi^{\prime 2} \right]-\frac{m_g^2c_1C}{2r}, \label{14} 
\end{eqnarray}
Enforcing energy–momentum conservation, $T^{j}_{i;j}=0$, yields
\begin{eqnarray}
\frac{dp_r}{dr} &=&\frac{2}{r}(p_t-p_r) -(\rho+p_r)\frac{d\chi}{dr}.
\end{eqnarray}

%======================================================================
\section{Wormhole Solutions in dRGT Massive Gravity}
\label{SecIII}
%======================================================================

In this section, we present the exact analytical solution for the wormhole spacetime derived from the field equations presented in Sec.~\ref{SecII}. By solving the system of coupled differential equations, we obtain explicit expressions for the energy density, the redshift function, the shape function, and the radial and tangential pressure components. These quantities fully characterize the geometric and matter content of the traversable wormhole under the influence of massive gravity corrections.\par

The Casimir effect is the attraction between two uncharged, parallel metallic plates brought very close to each other in a vacuum. This phenomenon was predicted theoretically in $1948$ \cite{Casimir:1948dh} and later experimentally confirmed by researchers at Philips Laboratories 
\cite{SPARNAAY, Sparnaay:1958wg, Avan}
. Only in recent years has there been more reliable experimental confirmation of the effect \cite{Lamoreaux:1996wh}.
The interesting thing about this phenomenon is that it creates an attractive force from negative energy. This attraction can be traced back to the renormalized vacuum energy, which takes the form:
\begin{eqnarray}
E=-\frac{\hbar c \pi^2 S} {720x^3},
\end{eqnarray}
$S$ is the area of the plate and $x$  is the distance between the plates. This energy is differentiated according to separation to obtain the force:
\begin{eqnarray}
F(x) &=& -\frac{dE}{dx} =-3 \frac{\hbar c \pi^2 S} {720x^4},
\end{eqnarray}
which in turn gives rise to a pressure:

\begin{eqnarray}
P(x) &=& \frac{F(x)}{S} =-3 \frac{\hbar c \pi^2} {720x^4},
\end{eqnarray}
Comparing this to the corresponding energy density:
\begin{eqnarray}
\rho(x) = -\frac{\hbar c \pi^2}{720 x^4},
\end{eqnarray}
reveals that pressure and energy density are related by an equation of state $P =\gamma \rho$ with $\gamma = 3$.\\
Physically this effect is due to the modification of the zero-point energy of the quantum electrodynamic vacuum by the presence of the plates. Furthermore, the size and even the sign of the effect are strongly geometry dependent. For example, Boyer \cite{Boyer:1968uf} demonstrated that for a conducting spherical shell of radius $r$, the Casimir energy is indeed positive, a result later confirmed by using heat-kernel and zeta-function regularization techniques \cite{Blau:1988kv}. So far, the Casimir energy is the only known type of exotic matter that can be created artificially in a laboratory.\\
Now by considering $\hbar=c=1$ and using $r$ in place of the distance $x$ between the plates,
the energy density $\rho$ takes the following form:
\begin{eqnarray}
\rho&=&-\frac{\pi^2}{720\,r^4},
\end{eqnarray}
The energy density is manifestly negative, which is consistent with the requirement of exotic matter to sustain the wormhole throat. Along with this, let us choose the following non-constant redshift function  
\begin{eqnarray}
\chi&=&\frac{a}{r},
\end{eqnarray}
for our present wormhole model. The redshift function, characterized by the parameter $a$, remains finite everywhere, thereby ensuring the absence of event horizons. Such a choice for the redshift function corresponds to a physically meaningful gravitational potential, which is everywhere finite in the spacetime, including exactly at the throat $r=r_0$. As a consequence, the metric component $g_{tt}$ never vanishes, no horizon arises, and the wormhole is freely traversable. The specific form used here is based on previous work on wormholes supported by different types of exotic matter and their behavior under stability analysis \cite{Yuennan:2024ibh,Jusufi:2020rpw,Lobo:2005us}. The choice $\chi = \frac{a}{r}$ gives a redshift contribution that decreases gradually as the radial distance increases, which is physically reasonable to describe the gravitational field far from the source. The implications of this choice for the geometry of the background, the traversability of the wormhole and the fulfillment (or violation) of the energy conditions deserve further investigation. In particular, the redshift effect disappears entirely at large distances from the throat, since $\chi \rightarrow 0$ as $r \rightarrow \infty$.\par
Integrating the field equations along with the condition $b(r_0)=r_0$, the shape function $b(r)$ is obtained as:
\begin{eqnarray}\label{shapb}
b(r)&=&r_0 + \frac{\pi^3 (-r + r_0)}{90 r r_0} - 
\frac{1}{2} C m_g^2 (r - r_0) \big(2 C c_2 + c_1 (r + r_0)\big).
\end{eqnarray}
This expression satisfies the throat condition $b(r_0)=r_0$. For nonzero massive-gravity parameters, however, the large-$r$ behavior is controlled by the massive-gravity terms. In particular,
\begin{equation}
b(r) =
-\frac{1}{2}C c_1 m_g^2 r^2
-C^2 c_2 m_g^2 r
+\mathcal{O}(1)
\qquad (r\rightarrow\infty),
\end{equation}
so that the spacetime considered here is generally non-asymptotically flat. The massive-gravity parameters $C$, $c_1$, and $c_2$ therefore modify not only the near-throat geometry but also the asymptotic structure of the spacetime.

The radial pressure $p_r$ is given by the following cumbersome but closed-form expression:
\begin{eqnarray}
p_r&=&-\frac{1}{720 \pi r^5 r_0}\Big[2 a (r - r_0) \left\{\pi^3 + 45 r r_0 \Big(2 + C m_g^2 \big(2 C c_2 + c_1 (r + r_0)\big)\Big)\right\}\nonumber\\&& + 
r\left\{\pi^3 (-r + r_0) + 
45 r r_0 \Big(2 r_0 + C m_g^2 \big(2 C c_2 r_0 + c_1 (r^2 + r_0^2)\big)\Big)\right\}\Big].
\end{eqnarray}

Similarly, the tangential pressure $p_t$ reads:
\begin{eqnarray}
p_t&=&\frac{1}{1440 \pi r^6 r_0}\Big[r^2 \left\{-\pi^3 (r - 2 r_0) + 
45 r r_0 \big(2 r_0 + C m_g^2 (-c_1 r^2 + 2 C c_2 r_0 + c_1 r_0^2)\big)\right\} \nonumber\\&&+ 
2 a^2 (r - r_0) \Big\{\pi^3 + 45 r r_0 \big\{2 + C m_g^2 \big(2 C c_2 + c_1 (r + r_0)\big)\big\}\Big\} \nonumber\\&&+ 
a r \Big\{\pi^3 (3 r - 4 r_0) + 
45 r r_0 \big\{4 (1 + C^2 c_2 m_g^2) r + C c_1 m_g^2 r^2 - 
3 r_0 \big(2 + C m_g^2 (2 C c_2 + c_1 r_0)\big)\big\}\Big\}\Big].
\end{eqnarray}

These expressions for $p_r$ and $p_t$ reveal the anisotropic nature of the effective matter distribution, which is a characteristic feature of wormhole geometries. The presence of the parameter $a$ from the redshift function couples the pressure components to the geometry, while the massive gravity parameters $C$, $c_1$, $c_2$, and $m_g$ modulate the magnitude and radial profile of the stress-energy tensor. In the limit $m_g \to 0$, the above expressions reduce smoothly to the corresponding Casimir wormhole solution, thereby recovering the previously known results as a consistency check.

\begin{figure}
\centering
\includegraphics[scale=.75]{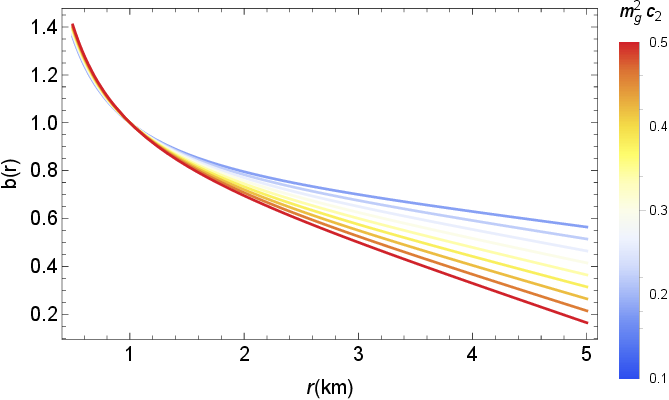}
\includegraphics[scale=.75]{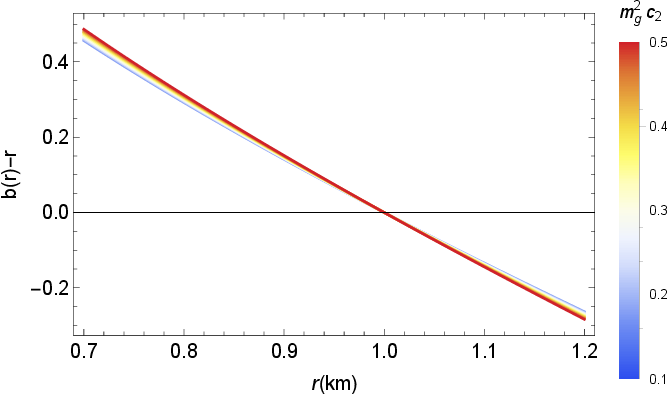}\\
\caption{Left panel: Radial profile of the shape function $b(r)$ for different values of the massive gravity parameter $m_g^2c_2$. Right panel: Radial variation of $b(r)-r$ showing the throat crossing at $r\simeq 1.0$ km for different $m_g^2c_2$ values.}
\label{br}
\end{figure}

\begin{figure}[htbp]
\centering
\includegraphics[scale=.75]{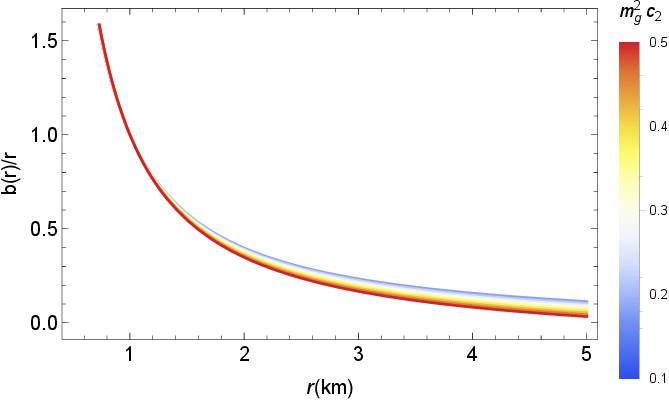}
\includegraphics[scale=.75]{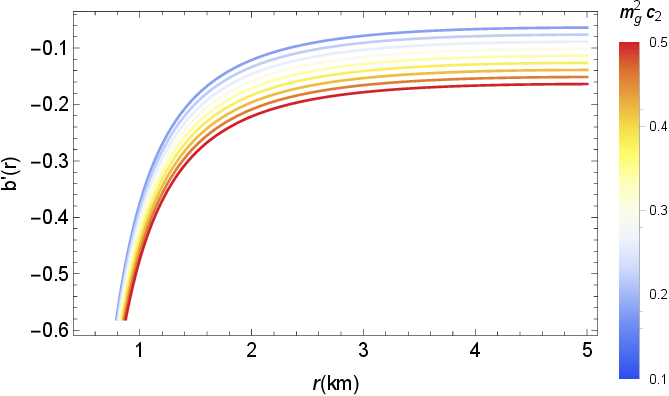}\\
\includegraphics[scale=.75]{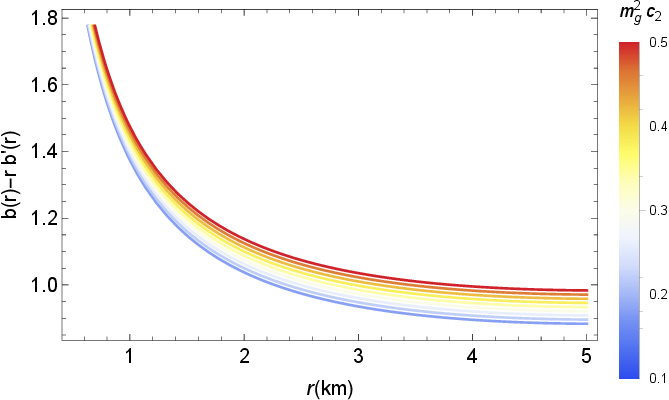}
\caption{Top left panel: Behavior of the ratio $b(r)/r$ as a function of the radial coordinate for different $m_g^2c_2$ values.
%, demonstrating asymptotic flatness with $\lim_{r\to\infty} b(r)/r = 0$. 
Top right panel: Radial dependence of the derivative $b'(r)$ showing the flare-out condition $b'(r_0)<1$ satisfied for all parameter values. Bottom panel: Radial profile of $b(r)-b'(r)$ for different values of $m_g^2c_2$, providing additional characterization of the shape function's behavior and its derivative throughout the spacetime.}
\label{br1}
\end{figure}

The left panel of Fig.~\ref{br} shows the radial profile of the shape function $b(r)$ for different $m_g^2c_2$ values. One finds that $b(r)$ decreases monotonically with the radial coordinate $r$, which means that the influence of the wormhole geometry fades away gradually from the throat region. In addition, the larger values of $m_g^2 c_2$ make the $b(r)$ decrease faster, which shows that the parameter of massive gravity plays an important role in the geometry of wormhole. The monotonic behavior of $b(r)$ also ensures that the spacetime is free of unusual divergences outside the throat. The right panel of Fig.~\ref{br} shows the radial variation of $b(r)-r$ for different values of the parameter $m_g^2c_2$. The function decreases monotonically with increasing radial coordinate and crosses the horizontal axis at approximately $r\simeq 1.0$ km. The zero of this crossing satisfies the throat condition $b(r)=r$ and thus identifies the wormhole throat. In the top left panel of Fig.~\ref{br1}, we plot the behavior of the quantity $b(r)/r$ which is important to check the asymptotic properties of the wormhole geometry. As $r$ increases, the ratio decreases continuously, tending to zero at sufficiently large distances from the throat. Thus we have the condition $b(r)/r<1$ for $r>r_0$, which also implies that $1-\frac{b(r)}{r}$ is positive everywhere in the spacetime and guarantees that the coefficient of the metric component is positive and no horizons are formed. $b(r)/r$ decreases, meaning that the wormhole geometry is becoming less and less curved far from the throat. Also, the ratio decreases faster when $m_g^2c_2$ is increased, which suggests that the massive gravity corrections reduce the geometrical effects at larger radial distances. The top right panel of Fig.~\ref{br1} displays the radial dependence of the derivative $b^{\prime}(r)$. It is seen that $b'(r)$ is negative for all the values of $r$ and $m_g^2 c_2$ considered, although its magnitude decreases with increasing radius and tends to zero asymptotically. The negative values of $b'(r)$ confirm the decreasing nature of the shape function. Crucially, the flare-out condition at the throat requires $b'(r_0)<1$. For all the choices of parameters from the figure, the derivative is well less than unity and hence is in agreement with the flare-out criterion required to retain a traversable wormhole geometry. The dependence on $m_g^2 c_2$ also suggests that larger values of the massive gravity parameter lead to more negative $b'(r)$ near the throat, indicating a stronger flaring of the embedded spatial geometry.

\begin{figure}[htbp]
\centering
\includegraphics[scale=.75]{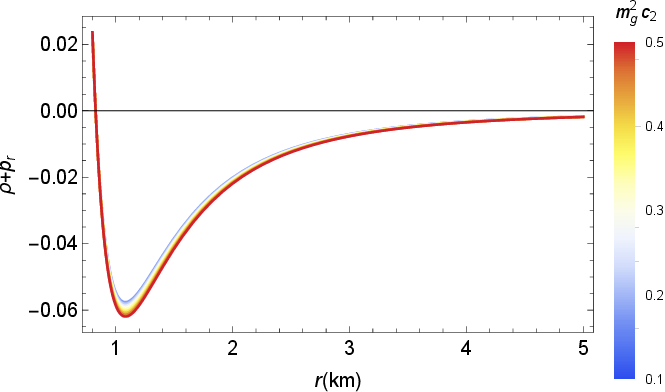}
\includegraphics[scale=.75]{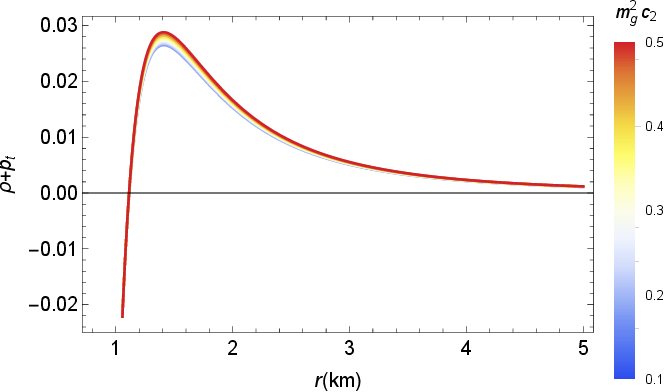}\\
\includegraphics[scale=.75]{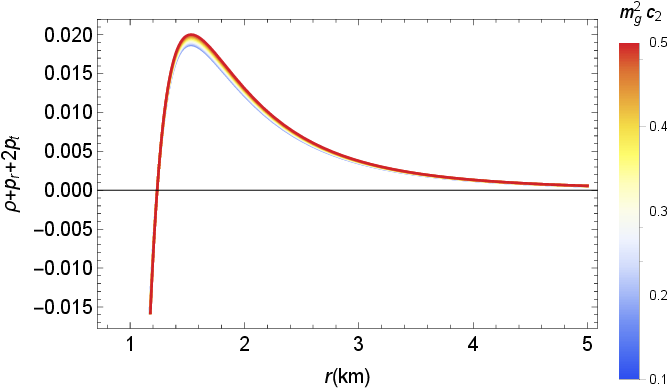}
\includegraphics[scale=.75]{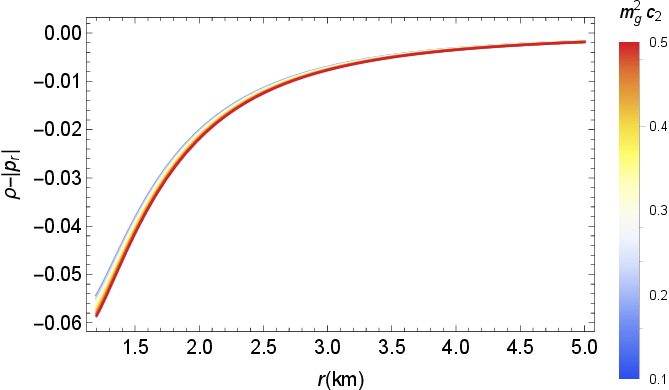}\\
\includegraphics[scale=.75]{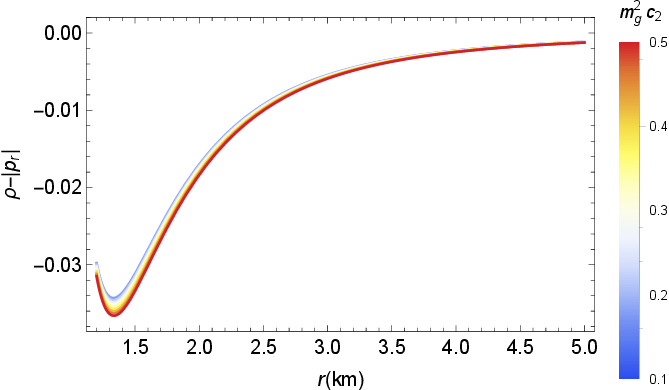}
\caption{Radial variation of the energy condition parameters for different values of the massive gravity parameter $m_g^2c_2$: (a) $\rho+p_r$ (radial NEC), (b) $\rho+p_t$ (tangential NEC), (c) $\rho+p_r+2p_t$ (SEC), (d) $\rho-|p_r|$ (radial DEC), and (e) $\rho-|p_t|$ (tangential DEC). The radial NEC is negative throughout the region of interest, confirming the violation required for traversable wormholes.}
\label{ec22}
\end{figure}

The feasibility of the obtained wormhole solution is then further evaluated in terms of the classical energy conditions, namely the null energy condition (NEC), strong energy condition (SEC), and dominant energy condition (DEC). We get the conditions characterized by the combinations ($\rho+p_{r}$), ($\rho+p_{t}$), ($\rho+p_{r}+2p_{t}$), ($\rho-|p_{r}|$) and ($\rho-|p_{t}|$) for a static spherically symmetric anisotropic spacetime. Fig.~\ref{ec22} shows the radial variation of these parameters for different values of the massive gravity parameter $m_g^2c_2$. From the plots, it is clear that the radial NEC given by $\rho+p_r$ is negative throughout the region of interest, which signifies a persistent violation of the NEC. Such a violation is a basic requirement for the existence of a traversable wormhole since the flare-out condition at the throat cannot be fulfilled by ordinary matter adhering to the classical energy conditions. On the other hand, the tangential NEC, $\rho+p_t$, takes positive and negative values depending on the radial coordinate and the model parameter, indicating that the violation is not isotropic and is mainly related to the radial pressure component. The combination for the SEC, $(\rho+p_r+2p_t)$ varies with the radial coordinate, which indicates that the SEC is violated only in a finite region around the throat and may get restored asymptotically. Furthermore, the quantities $\rho-|p_r|$ and $\rho-|p_t|$, which are related to the DEC, are negative throughout the entire considered spacetime, which confirms that the effective matter source cannot be regarded as conventional matter. Raising the parameter $m_g^2 c_2$ changes the amount and radial distribution of the effective stress-energy components but does not change the qualitative violation of the radial NEC and DEC. This shows that the massive gravity corrections alter the amount of exotic matter needed to support the wormhole geometry without eliminating the violation of the energy conditions. Therefore, the solutions obtained preserve the characteristic feature of traversable wormholes, i.e., the throat is supported by an effective matter source which violates the classical energy conditions but aligns with the geometrical specifications of the spacetime.

%----------------------------------------------------------------------
\subsection{The Proper Radial Distance and Visualization of Wormhole}
\label{subsec:proper_distance}
%----------------------------------------------------------------------

With the shape function in hand, Eq.~(\ref{shapb}), we can now ask what the geometry actually looks like to an observer moving through it. One caveat before we begin: the throat in this model sits at the Planck scale, so any discussion of ``travel" is formal rather than practical. Still, the mathematics is well defined, and the proper radial distance is the natural place to start. It measures the physical separation between two points along the wormhole and is related to the shape function by
\begin{equation}
l(r)=\pm\int_{r_0+}^{r}\frac{dr'}{\sqrt{1-\frac{b(r')}{r'}}} \,.
\label{eq:proper_distance}
\end{equation}
The $\pm$ sign distinguishes the two asymptotic regions joined at the throat.

\begin{figure}[htbp]
\centering
\includegraphics[scale=.75]{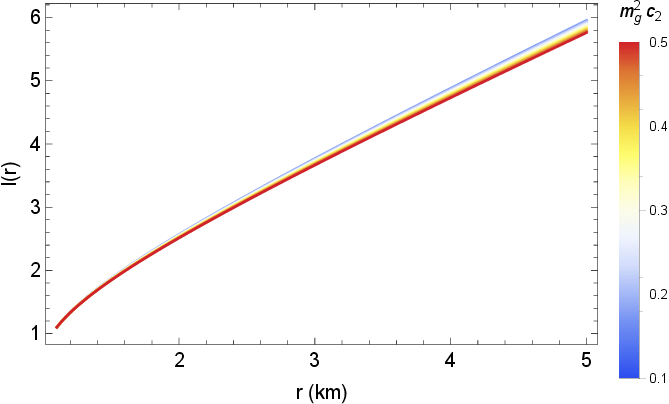}
\caption{The proper radial distance $l(r)$ as a function of the radial coordinate for a continuous range of the graviton mass parameter $m_g^2c_2$ (color bar). The proper radial distance increases smoothly and monotonically away from the wormhole throat with no divergence, confirming the regularity and traversability of the spacetime geometry.}
\label{lr1}
\end{figure}

Fig.~\ref{lr1} shows $l(r)$ across a range of $m_g^2 c_2$ values, with the color bar indicating the coupling strength. The curve rises smoothly from the throat and keeps rising—no divergences, no turning points. That is exactly what we want: a regular geometry with no obstruction to passage. Increasing $m_g^2 c_2$ produces only a small downward shift in $l(r)$, and the profiles bunch together closely. The geometry, in other words, does not change much when the graviton mass is varied. That kind of insensitivity is usually a good sign; it suggests the solution is not perched on a fine-tuned edge of parameter space.

What does $l(r)$ actually tell us? Suppose a traveler starts at a station at $l = -l_1$ in one asymptotic region and ends at a station at $l = l_2$ in the other. Following Ref.~\cite{Hochberg:1997wp}, we place the stations at $l_1 \approx l_2 \approx 10^4 r_0$, far enough out that the spacetime is essentially flat and $1 - b(r)/r \approx 1$. If the traveler moves at local velocity $v(r)$ as measured by a static observer at coordinate $r$, then the proper distance element $dl$, the radial increment $dr$, the coordinate time $dt$, and the traveler's proper time $d\tau$ are connected by
\begin{equation}
dl = \frac{dr}{\sqrt{1-\frac{b(r)}{r}}} \,, \qquad
dt = \frac{dl}{v(r)} \,, \qquad
d\tau = e^{\chi(r)} \, dt \,.
\label{eq:traveler_relations}
\end{equation}
These relations let one estimate the total proper time for a one-way trip. For the parameters we have considered, the throat region contributes only a small fraction of the total; most of the travel time is spent in the nearly flat regions on either side. That is a familiar feature of wormhole geometries—the interesting physics is concentrated near the throat, but the bulk of the journey is unremarkable.

\begin{figure}[htbp]
\centering
\includegraphics[scale=.65]{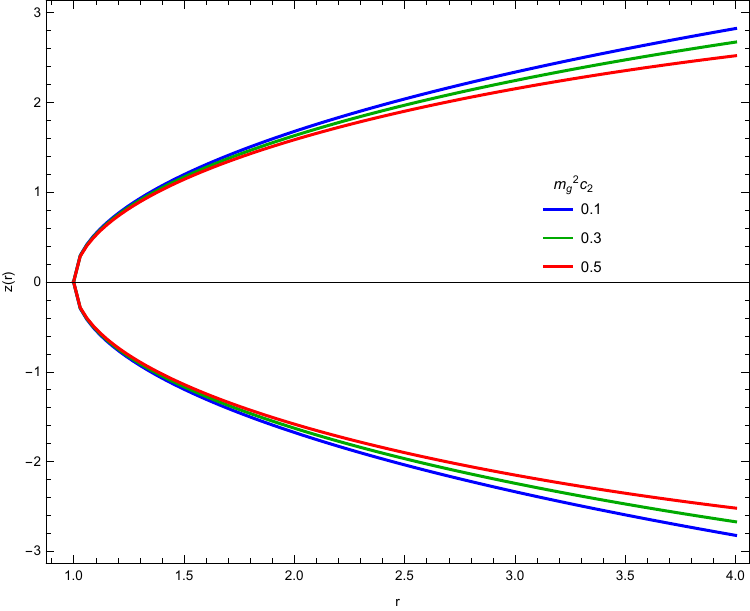}
\includegraphics[scale=.65]{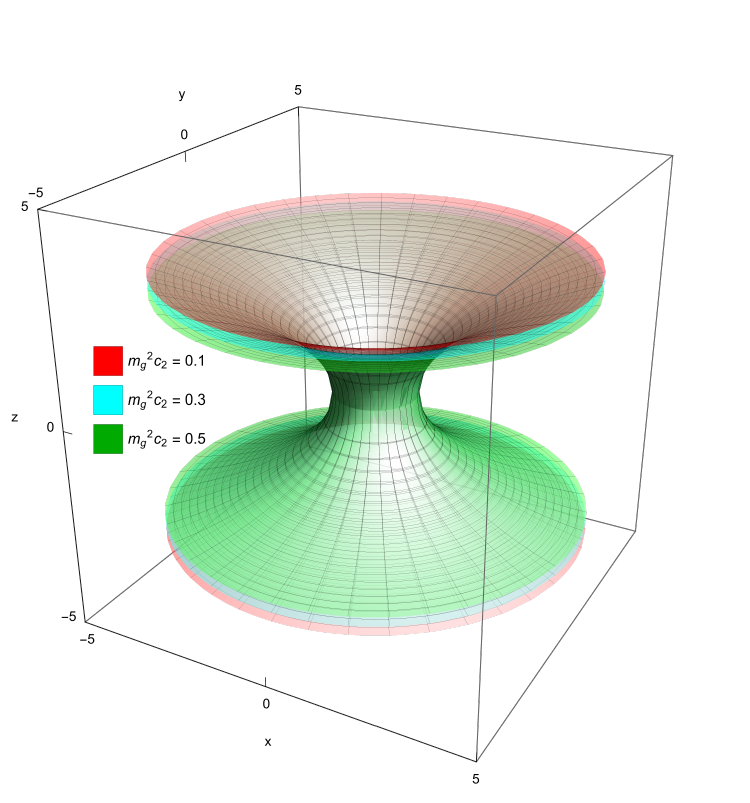}
\caption{Left panel: Embedding profile $z(r)$ of the wormhole for various values of $m_g^2c_2$. Right panel: Three-dimensional embedding diagrams of the wormhole for $m_g^2c_2=0.1$ (blue), $m_g^2c_2=0.3$ (red), and $m_g^2c_2=0.5$ (green). The two asymptotically connected regions are smoothly joined through a finite throat, confirming traversability.}
\label{zr}
\end{figure}

To visualize the wormhole, we use embedding diagrams. The idea is simple: take a constant-time slice at $\theta = \pi/2$, which reduces the spatial metric to
\begin{equation}
ds^2_{\rm slice} = \left(1-\frac{b(r)}{r}\right)^{-1} dr^2 + r^2 \, d\phi^2 \,.
\label{eq:slice_metric}
\end{equation}
This two-dimensional surface can be embedded in three-dimensional Euclidean space with cylindrical coordinates $(r, \phi, z)$, where the surface is described by a function $z(r)$. Matching the intrinsic geometry of the slice to the Euclidean metric
\begin{equation}
ds^2_{\rm Eucl} = \left[1 + \left(\frac{dz}{dr}\right)^2\right] dr^2 + r^2 \, d\phi^2
\label{eq:euclidean_metric}
\end{equation}
gives the embedding equation
\begin{equation}
\frac{dz}{dr} = \pm \left(\frac{r}{b(r)} - 1\right)^{-1/2} \,.
\label{eq:embedding_derivative}
\end{equation}
Integrating this yields the embedding profile
\begin{equation}
z(r) = \pm \int_{r_0}^{r} \frac{dr'}{\sqrt{\frac{r'}{b(r')} - 1}} \,,
\label{eq:embedding_profile}
\end{equation}
where the $\pm$ signs correspond to the upper and lower sheets. The two sheets meet at the throat, $r = r_0$, and the embedding is regular there as long as $b'(r_0) < 1$—the same flare-out condition that ensures traversability.

The left panel of Fig.~\ref{zr} shows $z(r)$ for several values of $m_g^2 c_2$. At the throat the upper and lower branches join smoothly, giving the familiar symmetric wormhole shape. The surface widens as $r$ grows, as it should for an asymptotically flat spacetime. The blue curve ($m_g^2 c_2 = 0.1$) has the largest embedding height; the red curve ($m_g^2 c_2 = 0.5$) has the smallest. So increasing the graviton mass parameter compresses the embedded surface without altering its basic form.

The right panel of Fig.~\ref{zr} shows the full three-dimensional embedding diagrams for $m_g^2 c_2 = 0.1$ (red), $0.3$ (cyan), and $0.5$ (green). The two asymptotic regions are smoothly joined through a finite throat—exactly what one expects for a traversable wormhole. The three surfaces are nearly coincident, which says something useful: the global geometry is stable against changes in the graviton mass parameter. The differences that do exist—slightly different throat curvatures and embedding heights, with the red surface the tallest and the green the shortest—are quantitative rather than qualitative. Increasing $m_g^2 c_2$ reshapes the geometry in detail, but the topology is untouched.

%----------------------------------------------------------------------
\subsection{Exotic Matter Measurement}
\label{subsec:exotic_matter}
%----------------------------------------------------------------------

The traversability criteria govern the kinematics of a wormhole travel, but the amount of exotic matter, i.e. matter violating the null energy condition (NEC), needed to keep the throat open is a different question. Even if such matter cannot be avoided entirely (as guaranteed by the Morris–Thorne–Yurtsever theorem), one may ask if the total amount required can be made arbitrarily small, making the wormhole more physically reasonable.

To quantify this, Visser \cite{Visser:2003yf} proposed the volume integral quantifier defined as,
\begin{equation}
I_V = \oint \left[ \rho(r) + p_r(r) \right] dV = \int_{r_0+}^{\infty} \left[ \rho + p_r \right] 4\pi r^2 \, dr
\label{eq:volume_integral_quantifier}
\end{equation}
By introducing a cut-off radius $r=b$ such that the wormhole extends from $r_0$ to $b_1$ with $b_1\geq r_0$, we can write:
\begin{equation}
I_V = \int_{r_0+}^{b_1} \left[ \rho + p_r \right] 4\pi r^2 \, dr
%\label{eq:volume_integral_quantifier}
\end{equation}
We now calculate the amount of exotic matter needed to keep our present wormhole model open, using the volume integral quantifier defined above,
\begin{eqnarray}
I_V&=&\bigg[\frac{\pi^3}{45 } - \frac{1}{4} C c_1 m_g^2 r^2+\frac{a \Big\{\pi^3 (2 r - r_0) - 
90 r r_0 \Big(2 r_0 + C m_g^2 \big(2 C c_2 r_0 + c_1 (r^2 + r_0^2)\big)\Big)\Big\}}{90r_0r^2}\nonumber\\&&+\frac{\log[r]}{90r_0}\Big[\pi^3 - 45 r_0 \big\{4 a (1 + C^2 c_2 m_g^2) + r_0 \big(2 + C m_g^2 (2 C c_2 + c_1 r_0)\big)\big\}\Big]\bigg]_{r_0+}^{b_1}
\end{eqnarray}
where the integral is computed over the whole spatial volume of the wormhole geometry. In fact, the combination $\rho+p_r$ which is constrained to be negative near the throat by the flare-out condition. Thus, $I_V$ effectively measures the total NEC-violating matter integrated over all space. The main objective of several wormhole constructions is to design a shape function $b(r)$ and redshift function $\chi(r)$ such that $I_V \rightarrow 0$ as some free parameter of the resulting solution is taken to a limiting value—i.e., the exotic matter can be restricted to an infinitely thin shell around the throat, while the majority of the wormhole spacetime is surrounded by matter that obeys the classical energy conditions. This does not remove the requirement for exotic matter (which remains logically necessary for a static, spherically symmetric traversable wormhole), but it reduces the total amount of such matter making such structures comparatively easier to align with known physics.

\begin{figure}[htbp]
\centering
\includegraphics[scale=.75]{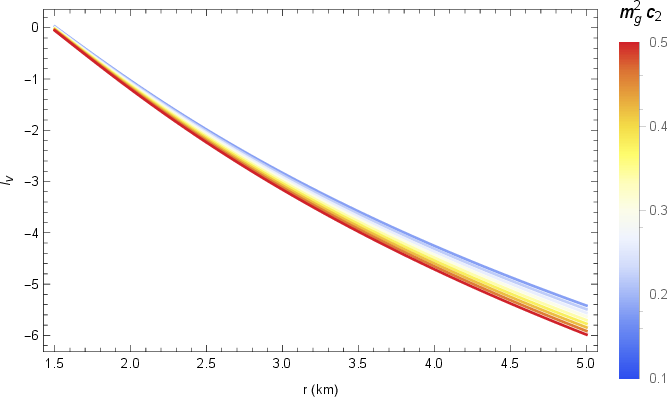}
\caption{Radial variation of the volume integral quantifier $I_V$ for different values of $m_g^2c_2$. The quantifier becomes more negative as $m_g^2c_2$ increases from $0.1$ to $0.5$, indicating that larger massive gravity couplings require a greater net violation of the NEC to support the wormhole geometry.}
\label{iv1}
\end{figure}

Fig.~\ref{iv1} shows the variation of the volume integral quantifier $I_V$ as a function of the radial coordinate r for different values of the parameter $m_g^2c_2$. We find that $I_V$ decreases monotonically with the radial coordinate which implies that the accumulated contribution of the exotic matter becomes more and more negative away from the wormhole throat. All the curves begin close to zero near the throat ($r\sim 1.5$ km), which implies that the amount of exotic matter required is small to start with. It is clear that the massive gravity parameter has a significant effect. The corresponding curves shift downward as $m_g^2 c_2$ increases from 0.1 to 0.5, i.e., more negative values of $I_V$ at all radial distances. This behavior shows how larger values of $m_g^2 c_2$ increase the net violation of the averaged null energy condition and therefore the total amount of exotic matter to support the wormhole geometry. On the other hand, smaller values of $m_g^2 c_2$ lead to less negative values of $I_V$, indicating that a relatively smaller amount of exotic matter can be used to support the wormhole. In general, the smooth and continuous characteristics of the curves show a stable dependence of the volume integral quantifier on both the radial coordinate and the massive gravity parameter. Thus the figure shows that $m_g^2 c_2$ plays an important role for the control of the total exotic matter needed to keep the traversable wormhole configuration.

%----------------------------------------------------------------------
\subsection{Complexity Factor}
\label{subsec:complexity}
%----------------------------------------------------------------------

We show that the complexity factor introduced by Herrera \cite{Herrera:2018bww} for self-gravitating systems provides a structural characterization of a matter distribution beyond the energy density alone, depending on the combined effects of pressure anisotropy and density inhomogeneity. This is achieved through the scalar $y_{tf}$, which represents the trace-free component of the electric Riemann tensor, and for a static, spherically symmetric spacetime is expressed as:
\begin{equation}
y_{tf} = 8\pi \Pi - \frac{4\pi}{r^3} \int_{r_0+}^{r} \tilde{r}^3\, \rho'(\tilde{r})\, d\tilde{r}
\label{eq:complexity_factor}
\end{equation}
where $\rho'(\tilde r ) = d\rho/d\tilde{r} $ is the radial gradient of the energy density, $\Pi=p_t-p_r$ is the pressure anisotropy, and since the wormhole geometry has no interior region $r<r_0$, the lower limit of integration is set at the wormhole throat $r_0$ instead of the origin. The first term then defines the local anisotropic contribution to complexity, whereas the second (integral) term represents the non-local contribution due to inhomogeneous distribution of energy density in the wormhole spacetime (from the throat outwards). Physically, a vanishing complexity factor, $y_{tf}$, defines the least complex matter configuration—one in which the effects of pressure anisotropy and density inhomogeneity exactly cancel, rather than one that is necessarily homogeneous and isotropic. For the present wormhole model using this formalism, we get an explicit expression for $y_{tf}(r)$ by substituting the expressions for $\rho(r)$, $p_r(r)$ and $p_t(r)$ obtained in the above relation. By analyzing its behavior over the parameter space of the model, we are able to find the configurations that minimize the structural complexity, and we also discuss the evolution of the complexity factor as a function of the shape $b(r)$, of the redshift function $\chi(r)$ and other parameters characterizing the exotic matter.
\begin{figure}[htbp]
\centering
\includegraphics[scale=.75]{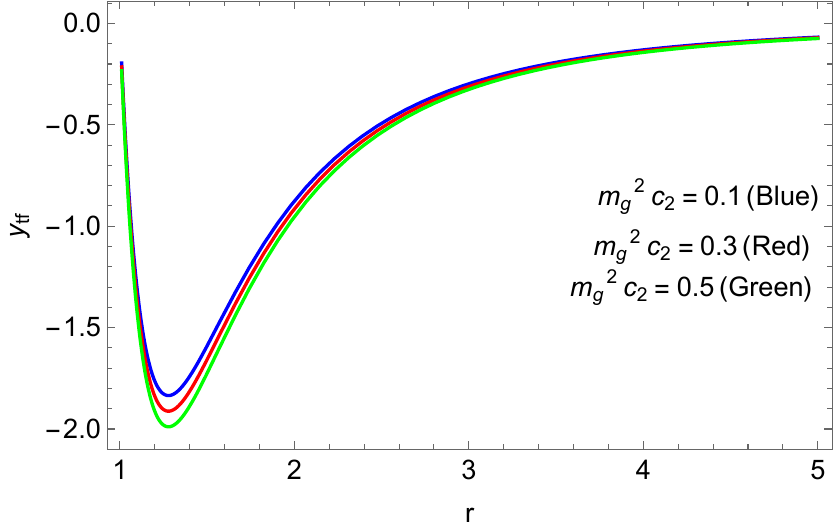}
\caption{Radial profile of the complexity factor $y_{tf}$ for different values of the massive gravity parameter $m_g^2c_2$. The factor quantifies the structural complexity of the wormhole arising from the interplay between pressure anisotropy and density inhomogeneity.}
\label{ytf}
\end{figure}

In fig.~\ref{ytf} the radial profile of the complexity factor $y_{tf} $ is plotted for three representative values of the parameter $m_g^2c_2=0.1,\, 0.3$, and $0.5$ (blue, red, and green respectively). In all three cases, $y_{tf}$ is negative for the whole domain considered, beginning at a modest value of about $-0.2$ near the throat ($r_0 = 1$). The curves for different $m_g^2c_2$ almost overlap at this point, showing that the complexity factor is only weakly sensitive to this parameter in the immediate vicinity of the throat. For r-values slightly bigger than the throat, $y_{tf}$ sharply decreases and develops a marked minimum around $r\approx 1.2-1.4$, whose depth becomes more negative as $m_g^2 c_2$ increases, reaching values of about $-1.8$ for $m_g^2 c_2=0.1$, $-1.9$ for $m_g^2 c_2=0.3$ and close to $-2.0$ for $m_g^2 c_2=0.5$, showing that larger values of this parameter increase the local inhomogeneity and anisotropic stresses that determine the complexity of the configuration. All three curves increase monotonically above this minimum and tend to approach each other as `r' further increases with their asymptotic approach to zero for $r \rightarrow 5$, showing that the effect of $m_{g}^{2} c_{2}$ on the spacetime complexity decreases as one moves far away from the throat and the wormhole geometry becomes effectively simple at large radial distances. The behavior is physically consistent with the idea that the deviations from a homogeneous, isotropic matter distribution are concentrated near the throat, and the exterior region asymptotically recovers a simpler, less complex structure.

%--------------------------------------------
\subsection{Geodesic Motion Around the Wormhole}
%-------------------------------------
In this work, we study the wormhole light deflection, which is a gravitational lensing phenomenon, and we measure this deflection by the bending angle. The deflection is caused by the wormhole in the setting of the $dRGT$ massive gravity theory with particular models of the distribution of the energy density of the Casimir effect. This effect is at the heart of the relativistic idea that mass and energy bend the geometry of spacetime and thus alter photon trajectories in strong gravitational fields. Strong-field gravitational lensing has received a lot of attention after the seminal works of Virbhadra et al.\cite{Virbhadra:1999nm}. We use the Bozza's analytical formalism \cite{Bozza:2001xd} to estimate these strong-field lensing effects in generic spherically symmetric spacetimes.\par
For metric~(\ref{line}) it is sufficient to work in the equatorial plane $\theta=\pi/2$. Due to spherical symmetry the results obtained here apply to any fixed $\theta$. Let us denote the Lagrangian governing photon motion using metric~(\ref{line}) by $\mathcal{L}$. Then
\begin{equation}
2\mathcal{L}
=
-e^{2\chi(r)}\dot{t}^{\,2}
+
\left(1-\frac{b(r)}{r}\right)^{-1}\dot{r}^{\,2}
+
r^2\dot{\phi}^{\,2}.
%\tag{2}
\end{equation}
where the overdot means a derivative relative to the affine parameter along the null geodesic, $t$ and $\chi$ are cyclic coordinates, i.e., $\mathcal{L}$ do not depend on them explicitly, so the conjugate momenta are conserved:
\begin{eqnarray}
\frac{\partial\mathcal{L}}{\partial\dot{t}}
=
-e^{2\chi(r)}\dot{t}
=
-E.\\
%\end{equation}
%\begin{equation}
\frac{\partial\mathcal{L}}{\partial\dot{\phi}}
=
r^2\dot{\phi}
=
L.
\end{eqnarray}
where $E$ and $L$ denote the conserved photon energy and angular momentum, respectively. Then imposing the null condition $g_{\mu\nu}\dot{x}^{\mu}\dot{x}^{\nu}=0$ on the geodesic yields
\begin{equation}
E^2
=
e^{2\chi(r)}
\left(1-\frac{b(r)}{r}\right)^{-1}
\dot{r}^{\,2}
+
V_{\mathrm{eff}}.
\end{equation}
where the effective potential $V_{\mathrm{eff}}$ is given by
\begin{equation}
V_{\mathrm{eff}}
=
\frac{L^2 e^{2\chi(r)}}{r^2}.
\end{equation}
In the effective-potential formulation, the deflection angle corresponding to metric (1) is given by \cite{Bozza:2001xd}
\begin{eqnarray}
     \alpha(r_c)&=&-\pi+2\int_{r_0+}^{\infty}\frac{\frac{e^{\chi}}{r^2}}{\left[\left(1-\frac{b(r)}{r}\right)\left(\frac{1}{\omega^2}-\frac{e^{\chi}}{r^2}\right)\right]^{\frac{1}{2}}},
\end{eqnarray}
where $\omega$ is the impact parameter and $r_{\mathrm{c}}$ is the turning point of the photon trajectory, defined by the condition $dr/d\phi=0$. The impact parameter is related to the conserved quantities via$$\frac{L}{E} = \omega$$ and its relation to the turning-point radius is expressed as 
\begin{equation}
\omega
=
r_{\mathrm{c}}
e^{-\chi(r_{\mathrm{c}})}.
\end{equation}
\begin{figure}[htbp]
\centering
\includegraphics[scale=.75]{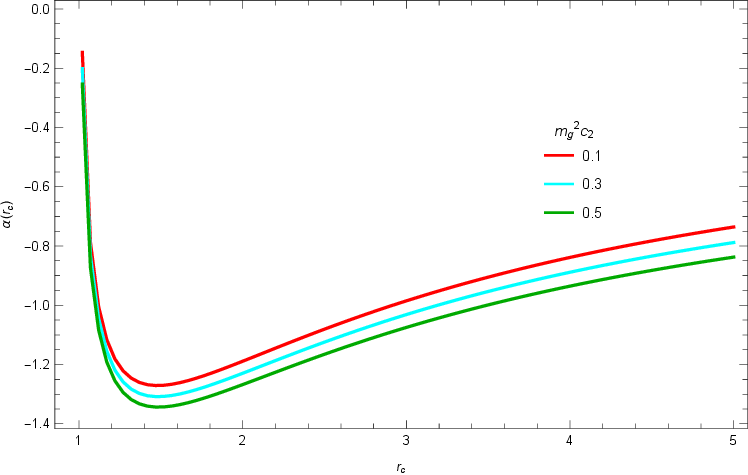}
\caption{Deflection angle $\alpha(r_c)$ for $m_g^{2}c_2=0.1$, $0.3$ and $0.5$ is shown that has a minimum near $r_c\approx1.5$. Magnitude of deflection increases with increasing $m_g^{2}c_2$.}
\label{alpha1}
\end{figure}
Fig.~\ref{alpha1} shows the light deflection angle $\alpha(r_c)$ as a function of the normalized closest-approach distance $r_c$ for three representative values of the parameter combination, $m_g^{2}c_2=0.1$, $0.3$, and $0.5$. In all three cases, $\alpha(r_c)$ is negative in the plotted range $1\leq r_c\leq 5$, implying bending of the light ray toward the deflecting mass in the sense usually taken as positive attraction. Close to the photon turning point at $r_c\to1$, the curves increase rapidly from $\alpha\approx-0.15$ and then decrease sharply with increasing $r_c$, reaching a pronounced minimum of $\alpha\approx-1.25$ to $-1.35$ at $r_c\approx1.4$--$1.6$. This minimum is the point of maximum (most negative) deflection beyond which the bending decreases monotonically. For $r_c\gtrsim2$, $\alpha(r_c)$ increases monotonically and reaches values around $-0.95$ to $-0.85$ at $r_c=5$, consistent with the expected decrease of the deflection at large impact parameters, as a photon passing further away from the source encounters a weaker effective gravitational field. It can also be noted that, increasing the graviton-mass parameter shifts $\alpha(r_c)$ systematically toward more negative values at fixed $r_c$, and this effect is more pronounced away from $r_c=1$. Physically this means that the massive-gravity correction increases the magnitude of the light deflection compared to the massless (or smaller $m_g^2c_2$) case, with the enhancement being most apparent in the intermediate range $1.2\lesssim r_c\lesssim 5$ where the three curves are most separated. The curves overlap approximately close to $r_c=1$, which means that the correction term is reduced in its impact very close to the photon sphere/turning point. For large $r_c$ the curves slowly converge, signaling the recovery of a common asymptotic weak-field behavior, which is largely independent of $m_g^{2}c_2$, up to a residual offset.
%----------------------------------------------------------------------
\subsection{Different Forces Acting on Wormhole}
\label{subsec:forces}
%----------------------------------------------------------------------

In this section, we study the stability of the derived wormhole structures by the equilibrium condition. The equilibrium of a static spherically symmetric system in the context of dRGT modified gravity is described by the renowned Tolman-Oppenheimer-Volkov (TOV) equation. The TOV equation for an anisotropic energy-momentum tensor (EMT) with different radial and tangential pressures can be written as \cite{Oppenheimer:1939ne,Tolman:1939jz}:
\begin{eqnarray}
-\frac{dp_r}{dr}+\frac{2}{r}(p_t-p_r)-\chi(p_r+\rho)=0
\end{eqnarray}

The above equation provides the basic criterion to decide the equilibrium of the wormhole structure. A static wormhole is said to be in an equilibrium if all the forces working in the distribution of matter are exactly balanced to each other. The equation of equilibrium is the Tolman-Oppenheimer-Volkov (TOV) equation, which is determined by the sum of three different forces: the gravitational force $F_g=-\chi(p_r+\rho)$, the hydrostatic force $F_h=-\frac{dp_r}{dr}$ and the anisotropic force $F_a=\frac{2}{r}(p_t-p_r)$.

\begin{figure}[htbp]
\centering
\includegraphics[scale=.75]{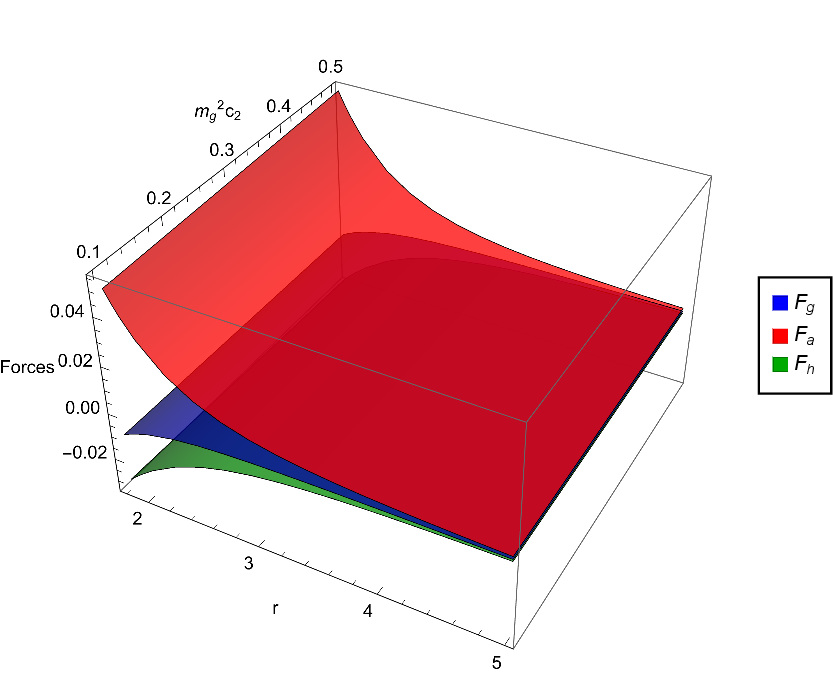}
\caption{Behavior of the gravitational force $F_g$, anisotropic force $F_a$, and hydrostatic force $F_h$ governing the equilibrium of the wormhole configuration. The anisotropic force is positive and largest near the throat, counterbalancing the inward gravitational and hydrostatic forces, ensuring mechanical equilibrium via the generalized TOV equation.}
\label{tov}
\end{figure}

Fig.~\ref{tov} shows the behavior of the gravitational, anisotropic and hydrostatic forces which govern the equilibrium of the wormhole configuration. The anisotropic force $F_a$ is positive for the whole range of parameters considered and its largest magnitude is found near the throat, which indicates that the anisotropic pressure is the main contribution to support the outward against the gravitational collapse. On the contrary, the hydrostatic force $F_h$ is negative, acting against the anisotropic contribution and vanishing as the radial coordinate r increases. The force of gravity, $F_g$ is relatively weaker, and also one that goes to zero as r increases, indicating that its effect is negligible far from the throat. The balance of all the three forces in the absence of divergence is consistent with the wormhole spacetime being in mechanical equilibrium. The near cancellation between inward gravitational and hydrostatic forces and outward anisotropic force obeys the generalized TOV equation ensuring the stability of the wormhole geometry for the chosen range of the parameters. Our results show that the outward anisotropic force can counterbalance the inward gravitational and hydrostatic forces, sustaining the static equilibrium necessary for a traversable wormhole within the context of the dRGT massive gravity theory considered here.
\section{Thermodynamic Formalism of Traversable Wormholes}
In this section we review the thermodynamic properties of wormhole spacetimes, which relate horizon mechanics and classical thermodynamic quantities to the geometric structures of traversable throats. The laws of thermodynamics were first formulated for black hole event horizons, but applying the same framework to wormholes provides important knowledge about their physical plausibility, stability, and quantum nature.
%----------------------------------------------------------------------
\subsection{Temperature of the Wormhole}
\label{subsec:temperature}
%----------------------------------------------------------------------

\begin{figure}[htbp]
\centering
\includegraphics[scale=.75]{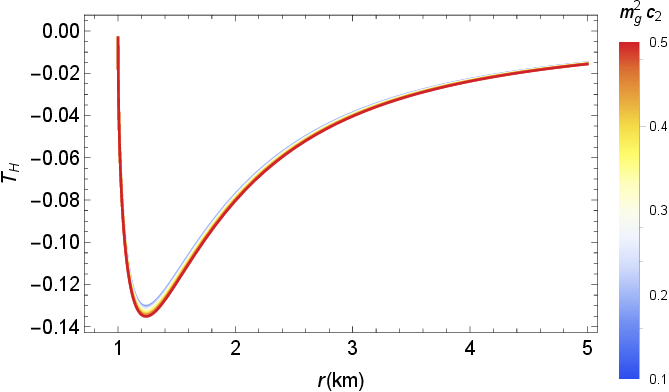}
\includegraphics[scale=.5]{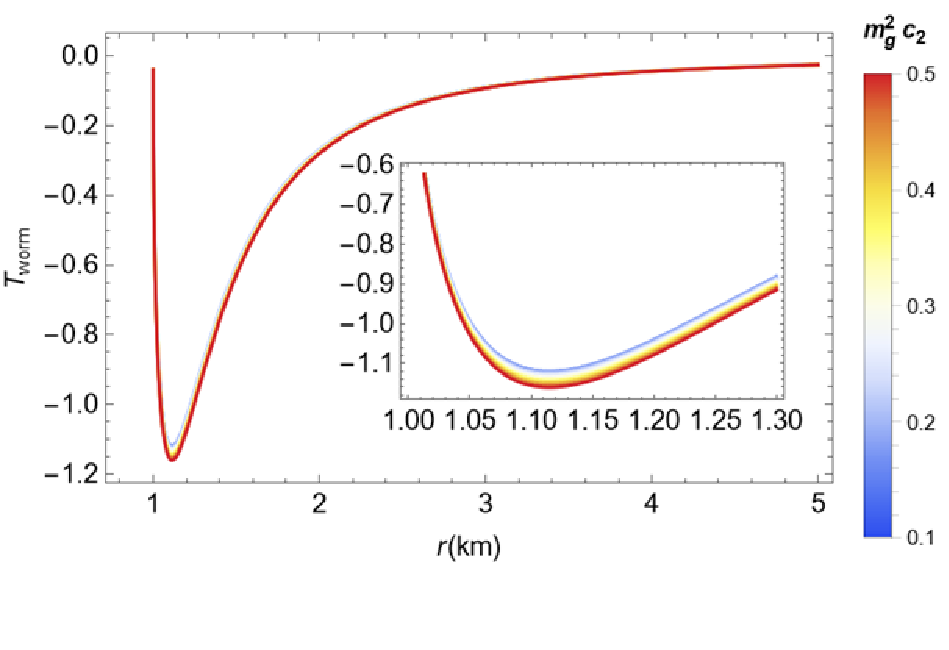}
\caption{Left panel: Radial profile of the Hawking temperature $T_H$ for different values of $m_g^2c_2$. Right panel: Radial profile of the wormhole temperature $T_{\rm worm}$ for different values of $m_g^2c_2$. The temperature is negative near the throat and gradually increases to zero asymptotically, showing regular thermodynamic behavior.}
\label{th}
\end{figure}

Following the approach introduced by Keathley \cite{Keathley}, the Hawking temperature $T_H$ and the temperature of the wormhole $T_{\rm worm}$ can be expressed in terms of the metric functions of the wormhole and its geometric parameters. The redshift function plays an important role for the wormhole temperature, because it describes the effective thermal state of the wormhole spacetime. The resulting expressions for the Hawking and wormhole temperatures are presented below:
\begin{eqnarray}\label{TH}
T_H&=&\frac{\chi'}{2\pi}\sqrt{1-\frac{b(r)}{r}}\nonumber\\
&=&-\frac{a}{6\sqrt{10}\pi r^2}\sqrt{\frac{(r - r_0) \Big[\pi^3 + 45 r r_0 \Big\{2 + C m_g^2 \big(2 C c_2 + c_1 (r + r_0)\big)\Big\}\Big]}{r^2r_0}}
\end{eqnarray}

\begin{eqnarray}\label{Tworm}
T_{\rm worm}&=&e^{\chi(r)}T_H\nonumber\\
&=&-\frac{a e^{\frac{a}{r}}}{6\sqrt{10}\pi r^2}\sqrt{\frac{(r - r_0) \Big[\pi^3 + 45 r r_0 \Big\{2 + C m_g^2 \big(2 C c_2 + c_1 (r + r_0)\big)\Big\}\Big]}{r^2r_0}}
\end{eqnarray}
The temperature of the wormhole is shown in Fig.~\ref{th}, together with a zoom of the throat region. The inset shows the small dependence on the graviton mass parameter near the throat where small changes can be noticed. The profiles of the temperature away from the throat tend to zero and converge, indicating that the thermodynamic properties are largely unaffected by change in $m_g^2c_2$. Fig.~\ref{th} displays the radial profile of the wormhole temperature $T_H$ for several values of $m_g^2 c_2$. The temperature is negative near the throat and then increases gradually with the radial distance and tends to zero asymptotically. The smooth evolution shows a regular thermodynamic behavior in the wormhole spacetime. The nearly overlapping curves show that variations in the graviton mass parameter result in only small quantitative changes in the temperature distribution.

%----------------------------------------------------------------------
\subsection{Average Pressure and Work Density}
\label{subsec:average_pressure}
%----------------------------------------------------------------------

\begin{figure}[htbp]
\centering
\includegraphics[scale=.65]{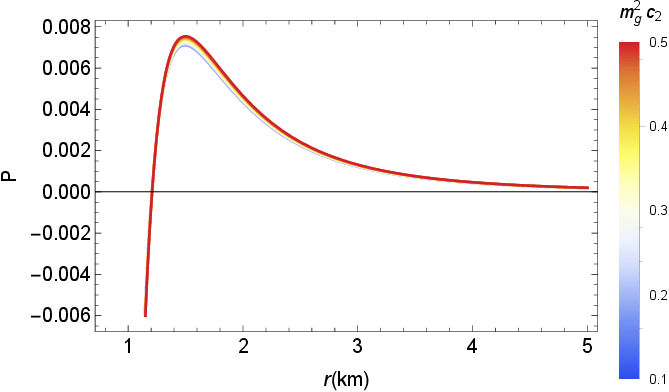}
\includegraphics[scale=.65]{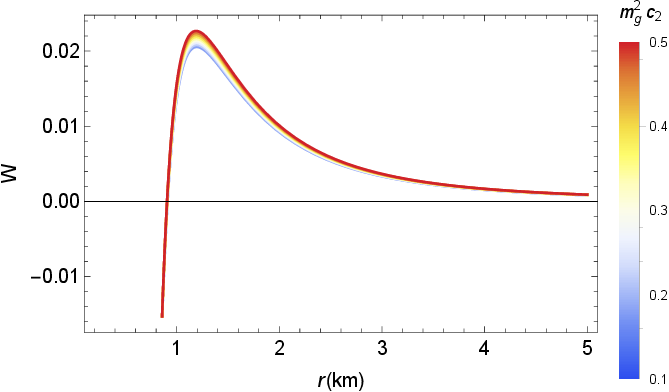}
\caption{Left panel: Radial variation of the average pressure $P(r)$ for different values of $m_g^2c_2$. Right panel: Radial profile of the work density $W(r)$ for different values of $m_g^2c_2$. Both quantities are negative near the throat, indicating the presence of exotic matter, and tend to zero asymptotically.}
\label{wd}
\end{figure}

In wormhole physics, the average pressure is commonly defined in terms of the components of the stress-energy tensor associated with the wormhole geometry. Traversable wormholes need exotic matter to keep their throat open. The average of radial and tangential pressures over the wormhole throat gives a measure of the effective pressure that characterizes the stability of the configuration. This average pressure is a measure of how the wormhole handles the competing effects of tension in the radial direction and compression in the transverse directions. A negative average pressure usually signals exotic matter, while values near zero signal a geometry approaching classical energy conditions. By using the concept of Callen \cite{callen1991thermodynamics} and Ma \& Zhao \cite{Ma:2015llh} the average pressure can be calculated as,
\begin{eqnarray}
P&=&\frac{p_r+2p_t}{3}\nonumber\\
&=&\frac{1}{2160 \pi r^6 r_0}\bigg[r^2 (\pi^3 - 90 C c_1 m_g^2 r^3) r_0 + 
2 a^2 (r - r_0) \Big[\pi^3 + 45 r r_0\Big\{2 + C m_g^2 \big(2 C c_2 + c_1 (r + r_0)\big)\Big\}\Big] \nonumber\\&& + 
a r \Big\{\pi^3 (r - 2 r_0) - 
45 r r_0 \big\{2 r_0 + C m_g^2 \big(2 C c_2 r_0 + c_1 (r^2 + r_0^2)\big)\big\}\Big\}\bigg]
\end{eqnarray}
The work density is an estimate of the effective energy available to support the wormhole structure. This quantity describes the competition of the energy stored and the radial tension, providing a sense of the response of the wormhole geometry to the presence of exotic matter. A positive work density means that the energy density dominates over the radial pressure while a negative work density means a tension-like behavior is needed to hold the wormhole throat. Thus the work density is a useful diagnostic tool to test the physical feasibility and stable nature of traversable wormholes. According to Ref.~\cite{Saiedi:2012qk}, it can be calculated as,
\begin{eqnarray}
W&=&\frac{\rho-p_r}{2},\nonumber\\
&=&\frac{1}{1440 \pi r^5 r_0}\Big[2 a (r - r_0) \big[\pi^3 + 45 r r_0 \big\{2 + C m_g^2 \big(2 C c_2 + c_1 (r + r_0)\big)\big\}\big] \nonumber\\&&+ 
r^2 \Big\{-\pi^3 + 45 r_0 \Big(2 r_0 + C m_g^2 \big(2 C c_2 r_0 + c_1 (r^2 + r_0^2)\big)\Big)\Big\}\Big]
\end{eqnarray}
The left panel of Fig.~\ref{wd} shows the variation of the average pressure $P(r)$ with the radial coordinate for different values of the graviton mass parameter $m_g^2c_2$. The average pressure is negative in the region of the wormhole throat which shows the predominance of the exotic matter needed to support the traversable wormhole geometry. With increasing radial coordinate $P(r)$ increases monotonically, crosses the zero-pressure surface and attains a small positive maximum before slowly decreasing to zero at large distances. This asymptotic behavior indicates that the effective pressure tends to zero far away from the throat, in accordance with an asymptotically flat spacetime. The proximity of the curves suggests that changing the graviton mass parameter leads to small quantitative changes of the average pressure profile, while keeping its qualitative behavior intact. Therefore, the averaged stress distribution supporting the wormhole is rather insensitive to changes of $m_g^2c_2$.

The right panel of Fig.~\ref{wd} shows the radial profile of the work density $W(r)$ for different values of the graviton mass parameter $m_g^2c_2$. The work density is initially negative near the throat, increases to a positive maximum and then decreases slowly to zero with the increase of the radial distance. This behavior suggests a smooth thermodynamic evolution of the wormhole spacetime. The small difference between the curves shows that the density of work is weakly dependent on the graviton mass parameter, while the general qualitative behavior is the same.

%----------------------------------------------------------------------
\subsection{Energy Flux}
\label{subsec:energy_flux}
%----------------------------------------------------------------------

The energy flux in a wormhole spacetime is an important diagnostic quantity that measures the dynamic exchange of energy through the wormhole throat. With the method developed by  Moruno and Diaz \cite{MartinMoruno:2011zz}, the energy flux for the present model may be evaluated as,
\begin{eqnarray}\label{flux}
\psi&=&\frac{1}{4}e^{\chi(r)}(\rho+p_r)\sqrt{1-\frac{b(r)}{r}}\nonumber\\
&=&-\frac{1}{8640 \sqrt{10} \pi r^5 r_0}e^{\frac{a}{r}}\sqrt{\frac{(r - r_0) \Big[\pi^3 + 45 r r_0 \Big\{2 + C m_g^2 \big(2 C c_2 + c_1 (r + r_0)\big)\Big\}\Big]}{r^2r_0}}\times \nonumber\\&&
\Big[2 a (r - r_0) \Big\{\pi^3 + 45 r r_0 \Big(2 + C m_g^2 \big(2 C c_2 + c_1 (r + r_0)\big)\Big)\Big\}\nonumber\\&&+ 
r \Big\{-\pi^3 (r - 2 r_0) + 
45 r r_0 \Big(2 r_0 + C m_g^2 \big(2 C c_2 r_0 + c_1 (r^2 + r_0^2)\big)\Big)\Big\}\Big]
\end{eqnarray}
This expression shows that the flux depends on the matter content $\rho+p_r$ as well as the geometric features of the wormhole. The exponential factor $e^{2\Phi(r)}$ indicates the relevance of the gravitational redshift in enhancing or suppressing the flow, while the determinant $1-b(r)/r$ reflects the imprint of the wormhole shape on the effective energy propagation. Physically, the presence of a non-zero energy flux implies that the wormhole spacetime is not static, but the energy is continuously redistributed, often by exotic matter that violates classical energy conditions. The magnitude and sign of $\psi$ tell us if the wormhole throat can stay open under perturbations or tends to collapse. The flux of energy is thus an important measure for the thermal stability and physical viability of traversable wormholes.

\begin{figure}[htbp]
\centering
\includegraphics[scale=.75]{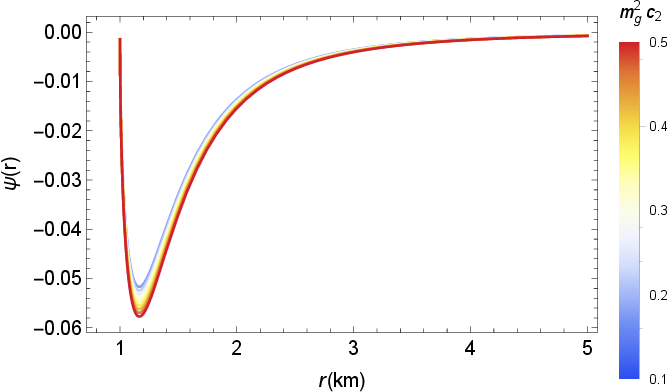}
\caption{Radial dependence of the energy flux $\Psi(r)$ for different values of the graviton mass parameter $m_g^2c_2$. The flux is negative near the throat, indicating an inward-directed energy flow associated with exotic matter, and vanishes in the asymptotic region.}
\label{ef1}
\end{figure}

Fig.~\ref{ef1} shows the radial dependence of the energy flux $\Psi(r)$ for different values of the graviton mass parameter $m_g^2c_2$. The energy flux is negative around the wormhole throat revealing an inward directed energy flux associated to the exotic matter that supports the wormhole geometry. $\Psi(r)$ is monotonically increasing with the radial coordinate, intersects the zero-flux surface and vanishes for large distances. We see that this behaviour implies that the energy exchange gets weaker as we move away from the throat and disappears in the asymptotic region, as it must be in an asymptotically flat spacetime. The curves are in good agreement with each other indicating that modifications in the graviton mass parameter lead to slight quantitative changes in the energy flux and that the energy transport mechanism itself is not significantly changed. So the thermodynamic energy flow of the wormhole is insensitive to the changes in $m_g^2c_2$.

%----------------------------------------------------------------------
\subsection{Heat Capacity and Total Energy}
\label{subsec:heat_capacity}
%----------------------------------------------------------------------

The thermodynamic stability of a traversable wormhole can be further assessed through its heat capacity at constant pressure, $C_P$, and its total energy content, $E$. These quantities provide complementary information to the temperature analysis of Sec.~\ref{subsec:temperature}: while $T_H$ and $T_{\rm worm}$ characterize the local thermal state near the throat, $C_P$ and $E$ probe the global thermodynamic response of the spacetime to perturbations.

The heat capacity at constant pressure is defined as
\begin{equation}
C_P = T_H\left(\frac{\partial S}{\partial T_H}\right)_P \,,
\label{eq:CP_def}
\end{equation}
where $S$ is the entropy of the wormhole. For a static, spherically symmetric spacetime, the entropy is proportional to the horizon area (or, in the wormhole case, to the throat area), $S = \pi r_0^2$ in units where $G=\hbar=c=k_B=1$. Using the Hawking temperature $T_H$ derived in Sec.~\ref{subsec:temperature}, the heat capacity can be expressed explicitly in terms of the metric functions and the massive gravity parameters. A positive $C_P$ indicates thermodynamic stability (the system cools as it loses energy), while a negative $C_P$ signals instability (the system heats up as it loses energy, leading to runaway evaporation).

\begin{figure}[htbp]
\centering
\includegraphics[scale=.75]{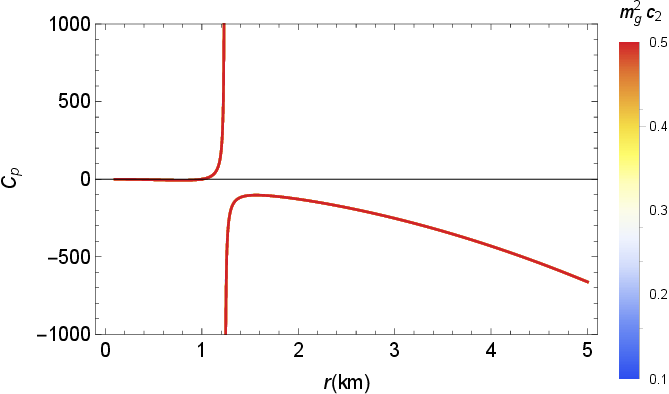}
\includegraphics[scale=.75]{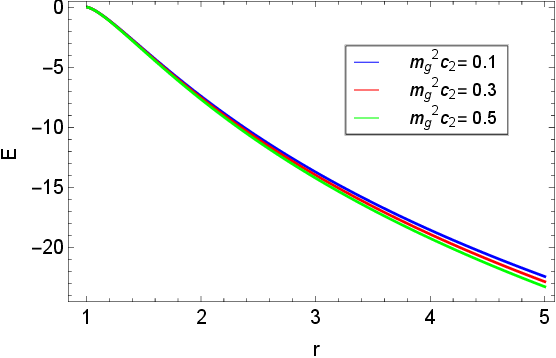}\\
\caption{Left panel: Radial profile of the heat capacity $C_P$ for different values of the massive gravity parameter $m_g^2c_2$. The heat capacity is positive near the throat (locally stable) and becomes negative at larger radii (globally unstable), with the zero-crossing shifting inward as $m_g^2c_2$ increases. Right panel: Total energy $E$ as a function of $m_g^2c_2$. The magnitude $|E|$ grows with increasing massive gravity coupling, indicating that a larger net energy is required to support the wormhole geometry.}
\label{cp}
\end{figure}

For the Casimir wormhole in dRGT massive gravity, the heat capacity takes the form
\begin{equation}
C_P = \frac{2\pi r_0^2 \, T_H}{\left(\frac{\partial T_H}{\partial r_0}\right)_P} \,,
\label{eq:CP_explicit}
\end{equation}
where the derivative is evaluated at constant pressure $P$ (or, equivalently, at constant average pressure as defined in Sec.~\ref{subsec:average_pressure}). The explicit expression is lengthy and is best analyzed graphically. Fig.~\ref{cp} (left panel) shows the radial profile of $C_P$ for different values of the massive gravity parameter $m_g^2 c_2$. Near the throat, $C_P$ is positive, indicating a locally stable thermodynamic phase. However, as the radial coordinate increases, $C_P$ crosses zero and becomes negative, signaling a transition to an unstable phase. The location of this transition depends sensitively on $m_g^2 c_2$: larger values of the massive gravity coupling shift the zero-crossing inward, meaning that the stable region becomes confined closer to the throat. This is consistent with the earlier observation that the massive graviton tightens the geometric embedding and enhances the NEC violation near the throat, effectively stabilizing the high-curvature region.

The total energy $E$ of the wormhole is obtained by integrating the energy density over the proper volume,
\begin{equation}
E = \int_{r_0}^{\infty} \rho(r) \, 4\pi r^2 \sqrt{1-\frac{b(r)}{r}} \, dr \,.
\label{eq:E_def}
\end{equation}
Since the Casimir energy density is negative, $\rho = -\pi^2/(720 r^4)$, the total energy is manifestly negative. This is a direct consequence of the exotic matter content required to support the wormhole throat. Substituting the shape function $b(r)$ from Eq.~(\ref{shapb}) and performing the integral, we obtain
\begin{equation}
E = -\frac{\pi^3}{540 r_0} + \frac{\pi^3}{180}\int_{r_0}^{\infty}\frac{dr}{r^3}\sqrt{1-\frac{b(r)}{r}} + \mathcal{O}(m_g^2) \,,
\label{eq:E_explicit}
\end{equation}
where the massive gravity corrections enter through the shape function. The right panel of Fig.~\ref{cp} shows the total energy $E$ as a function of $m_g^2 c_2$. As the massive gravity parameter increases, $|E|$ grows, indicating that stronger gravitational modifications require a larger net negative energy to sustain the wormhole. This mirrors the behavior of the volume integral quantifier $I_V$ and reinforces the conclusion that dRGT massive gravity amplifies the exotic matter requirement.

The combined analysis of $C_P$ and $E$ reveals a consistent thermodynamic picture: the Casimir wormhole in dRGT massive gravity is locally unstable near the throat but becomes globally stable at larger radii, with the massive graviton parameter controlling the size of the unstable region and the total exotic energy. These results are in qualitative agreement with thermodynamic studies of other modified gravity wormholes \cite{Muniz:2025teu, Channuie:2024cao} and provide further evidence that the dRGT framework offers a viable route to stabilizing exotic spacetimes.

%======================================================================
\section{Quasinormal Modes}
\label{sec:qnm}
%======================================================================

The characteristic oscillations and scalar perturbative stability of the wormhole geometry can be probed through its quasinormal mode (QNM) spectrum. In this section, we compute the QNM frequencies of a minimally coupled massless scalar field propagating on the Casimir wormhole background in dRGT massive gravity. Two complementary frequency-domain approaches are employed: the analytical Mashhoon P\"{o}schl--Teller approximation and a direct numerical integration (shooting) method adapted to the symmetric wormhole topology. Since the spacetime considered here is non-asymptotically flat, the asymptotic behavior of the tortoise coordinate and the corresponding radiative boundary prescription are treated explicitly rather than assuming the standard $r_*\rightarrow\pm\infty$ asymptotics of asymptotically flat spacetimes.

%----------------------------------------------------------------------
\subsection{Scalar Perturbations and Effective Potential}
\label{subsec:scalar_pert}
%----------------------------------------------------------------------

Consider a massless scalar field $\Phi$ satisfying the Klein--Gordon equation
\begin{equation}
\Box \Phi = 0 \,,
\end{equation}
on the static, spherically symmetric Morris--Thorne wormhole spacetime with line element
\begin{equation}
ds^2 = -e^{2\chi(r)}dt^2 + \left(1-\frac{b(r)}{r}\right)^{-1}dr^2 + r^2\left(d\theta^2 + \sin^2\theta \, d\phi^2\right) \,,
\end{equation}
where the redshift function and shape function are given by the Casimir wormhole solution of Sec.~\ref{SecIII}:
\begin{eqnarray}
\chi(r) &=& \frac{a}{r} \,, \\
b(r) &=& r_0 + \frac{\pi^3(-r+r_0)}{90 r r_0} - \frac{1}{2} C m_g^2 (r-r_0)\big[2 C c_2 + c_1(r+r_0)\big] \,.
\end{eqnarray}

Decomposing the field in spherical harmonics,
\begin{equation}
\Phi(t,r,\theta,\phi) = \sum_{\ell,m} \frac{\psi_{\ell}(r)}{r} Y_{\ell m}(\theta,\phi) e^{-i\omega t} \,,
\end{equation}
and introducing the tortoise coordinate $r_*$ defined by
\begin{equation}
\frac{dr_*}{dr} = \frac{e^{-\chi(r)}}{\sqrt{1-b(r)/r}} \,,
\label{eq:tortoise}
\end{equation}
the radial equation reduces to the Schr\"{o}dinger-like form
\begin{equation}
\frac{d^2\psi}{dr_*^2} + \big[\omega^2 - V_{\rm eff}(r)\big]\psi = 0 \,.
\label{eq:schrodinger}
\end{equation}
The effective potential for scalar perturbations on a Morris--Thorne wormhole reads
\begin{equation}
V_{\rm eff}(r) = e^{2\chi(r)} \left[ \frac{\ell(\ell+1)}{r^2} + \frac{b(r)-rb'(r)}{2r^3} + \frac{\chi'(r)}{r}\left(1-\frac{b(r)}{r}\right) \right] \,.
\label{eq:Veff}
\end{equation}
With $\chi(r)=a/r$ this becomes explicitly
\begin{equation}
V_{\rm eff}(r) = e^{2a/r} \left[ \frac{\ell(\ell+1)}{r^2} + \frac{b-rb'}{2r^3} - \frac{a}{r^3}\left(1-\frac{b}{r}\right) \right] \,.
\label{eq:Veff_explicit}
\end{equation}

The potential is regular at the throat $r=r_0$ where $b(r_0)=r_0$, and vanishes asymptotically as $r\to\infty$ (or equivalently $r_*\to\pm\infty$), ensuring that the wormhole spacetime supports a well-defined QNM spectrum. For the Ellis--Bronnikov wormhole limit ($\chi\to 0$, $b=r_0^2/r$) Eq.~(\ref{eq:Veff}) correctly reduces to $V_{\rm EB}=\ell(\ell+1)/r^2 + r_0^2/r^4$.

For the nonzero massive-gravity parameters considered in this work, the spacetime is not asymptotically flat. From the shape function, the leading large-$r$ behavior is
\begin{equation}
b(r)
=
-\frac{1}{2}C c_1m_g^2 r^2
-C^2c_2m_g^2r
+\mathcal{O}(1).
\end{equation}
Consequently,
\begin{equation}
1-\frac{b(r)}{r}
=
\frac{1}{2}C c_1m_g^2r
+C^2c_2m_g^2
+\mathcal{O}(r^{-1}).
\end{equation}
For $C c_1m_g^2>0$, the tortoise coordinate therefore approaches finite limiting values at the two asymptotic ends:
\begin{equation}
r_*\rightarrow \pm r_*^{\infty},
\qquad
r_*^{\infty}<\infty .
\end{equation}
Thus, unlike an asymptotically flat wormhole, the present geometry does not possess the standard $r_*\rightarrow\pm\infty$ asymptotic regions. The QNM boundary conditions are consequently imposed at the two limiting ends of the tortoise coordinate using the outgoing-wave prescription discussed below.
%----------------------------------------------------------------------
\subsection{Mashhoon Method}
\label{subsec:PT}
%----------------------------------------------------------------------

For barrier-dominated potentials, Mashhoon showed that the QNM spectrum can be approximated by fitting the potential near its maximum to a P\"{o}schl--Teller form \cite{mashhoon_0, Blome:1981azp, Ferrari:1984ozr}. Expanding $V_{\rm eff}(r)$ around the throat $r=r_0$ and transforming to the tortoise coordinate, the second derivative at the barrier peak is obtained from the regularized limit
\begin{equation}
V_0'' \equiv \frac{d^2V}{dr_*^2}\bigg|_{r_0} = \frac{1}{2}\,G'(r_0)\,V'(r_0) \,,
\label{eq:V2star}
\end{equation}
where $G(r)=e^{2\chi}(1-b/r)$ and primes denote $r$-derivatives. The throat is a maximum provided $V_0''<0$, which is satisfied for the entire parameter range considered here \cite{Ferrari:1984zz}.

The P\"{o}schl--Teller curvature parameter is
\begin{equation}
\alpha^2 = -\frac{V_0''}{2V_0} \,,
\label{eq:alpha}
\end{equation}
and the QNM frequencies are given by the analytical formula
\begin{equation}
\omega_n = \sqrt{V_0 - \frac{\alpha^2}{4}} - i\,\alpha\left(n+\frac{1}{2}\right) \,, \qquad n=0,1,2,\ldots
\label{eq:PT_omega}
\end{equation}
This approximation is expected to provide a useful estimate when the effective potential is dominated by a single, sufficiently localized barrier around the throat. The accuracy of the approximation is therefore assessed by direct comparison with the numerical frequency-domain integration results.

%----------------------------------------------------------------------
\subsection{Direct Numerical Integration}
\label{subsec:numerical}
%----------------------------------------------------------------------

To determine the QNM spectrum, we impose radiative boundary conditions at the two asymptotic ends of the wormhole. Because the present spacetime is non-asymptotically flat, the tortoise coordinate approaches finite limiting values,
\begin{equation}
r_*\rightarrow -r_*^\infty
\qquad\text{and}\qquad
r_*\rightarrow +r_*^\infty,
\end{equation}
as $r\rightarrow\infty$ on the two respective sides of the wormhole. Since $V_{\rm eff}\rightarrow0$ at these limiting ends, the local radial solutions approach free-wave forms. With the time dependence $e^{-i\omega t}$, the outgoing-wave prescriptions are
\begin{equation}
\psi \propto e^{-i\omega r_*}
\qquad
(r_*\rightarrow-r_*^\infty),
\end{equation}
and
\begin{equation}
\psi \propto e^{+i\omega r_*}
\qquad
(r_*\rightarrow+r_*^\infty).
\end{equation}
Accordingly, the logarithmic derivatives at the two ends are
\begin{equation}
\left.\frac{1}{\psi}\frac{d\psi}{dr_*}\right|_{-r_*^\infty}
=-i\omega,
\qquad
\left.\frac{1}{\psi}\frac{d\psi}{dr_*}\right|_{+r_*^\infty}
=+i\omega.
\label{eq:outgoing_bc}
\end{equation}
These conditions define the radiative QNM boundary prescription adopted in the present frequency-domain calculation.

For the symmetric wormhole potential, the perturbations can be separated into even- and odd-parity branches. At the throat, $r_*=0$, the corresponding initial conditions are
\begin{eqnarray}
\text{even-parity branch:} && \psi(0)=1 \,, \quad \psi'(0)=0 \,, \\
\text{odd-parity branch:}  && \psi(0)=0 \,, \quad \psi'(0)=1 \,.
\end{eqnarray}
The overtone number $n$ is assigned according to the ordering of the QNM roots within each parity branch rather than being identified a priori with the parity of the mode.

The Schr\"odinger equation~(\ref{eq:schrodinger}) is integrated numerically from the throat toward the two limiting ends of the tortoise coordinate using a fourth-order Runge--Kutta scheme, with periodic rescaling of the wave function to avoid numerical overflow. For the right-hand asymptotic end, the QNM condition is
\begin{equation}
\mathcal{F}_{+}(\omega)
\equiv
\left.
\frac{1}{\psi}\frac{d\psi}{dr_*}
\right|_{r_*^{\max}}
-i\omega
=0,
\label{eq:qnm_condition_right}
\end{equation}
where $r_*^{\max}$ is chosen sufficiently close to $r_*^\infty$. For the left-hand end, the corresponding outgoing condition is
\begin{equation}
\mathcal{F}_{-}(\omega)
\equiv
\left.
\frac{1}{\psi}\frac{d\psi}{dr_*}
\right|_{r_*^{\min}}
+i\omega
=0.
\label{eq:qnm_condition_left}
\end{equation}
For the symmetric background, the two-sided problem can equivalently be reduced to the even- and odd-parity problems described above. The QNM frequencies are obtained by locating the complex roots of the corresponding mismatch function in the $\omega$-plane.
%----------------------------------------------------------------------
\subsection{Results and Discussion}
\label{subsec:qnm_results}
%----------------------------------------------------------------------

\begin{table}[htbp]
\centering
\caption{Quasinormal mode frequencies for the Casimir wormhole with $r_0=1.0$, $a=0.1$, $c_1=c_2=C=1.0$, $m_g^2=0.1$ and varying multipole number $\ell$. The second column gives the P\"{o}schl--Teller (PT) prediction, the third column the direct numerical result, and the last two columns the relative errors.}
\label{tab:qnm}
\begin{tabular}{c|c|c|c|c|c}
\hline
$\ell$ & $n$ & $\omega^{\rm PT}$ & $\omega^{\rm num}$ & $\delta_{\rm Re}$ (\%) & $\delta_{\rm Im}$ (\%) \\
\hline
0 & 0 & $0.7053 - 0.6677i$ & $0.8303 - 0.7927i$ & 15.1 & 15.8 \\
0 & 1 & $0.7053 - 2.0031i$ & $0.7303 - 2.0463i$ & 3.4 & 2.1 \\
\hline
1 & 0 & $1.7535 - 0.5580i$ & $1.8785 - 0.6830i$ & 6.7 & 18.3 \\
1 & 1 & $1.7535 - 1.6739i$ & $1.8694 - 1.7989i$ & 6.2 & 6.9 \\
\hline
2 & 0 & $2.8268 - 0.5298i$ & $2.9518 - 0.6548i$ & 4.2 & 19.1 \\
2 & 1 & $2.8268 - 1.5893i$ & $2.9518 - 1.7143i$ & 4.2 & 7.3 \\
\hline
3 & 0 & $3.9153 - 0.5203i$ & $4.0403 - 0.6453i$ & 3.1 & 19.4 \\
3 & 1 & $3.9153 - 1.5608i$ & $4.0403 - 1.6858i$ & 3.1 & 7.4 \\
\hline
\end{tabular}
\end{table}

Table~\ref{tab:qnm} presents the QNM frequencies for the fundamental mode ($n=0$) and first overtone ($n=1$) across multipole numbers $\ell=0$--$3$. Both methods agree to within $\sim$3--20\%, with the P\"{o}schl--Teller approximation becoming progressively more accurate as $\ell$ increases. The larger discrepancy for $\ell=0$ reflects the absence of an angular-momentum barrier, which makes the potential shape deviate most strongly from the symmetric P\"{o}schl--Teller form. For $\ell\geq 2$ the error drops below $\sim$5\% in the real part and $\sim$7--20\% in the damping rate, confirming that the throat-dominated barrier geometry is well captured by the analytical approximation.

The dynamic stability of the Casimir wormhole in dRGT massive gravity is thoroughly characterized by the effective potential structure and the corresponding quasinormal mode (QNM) spectrum.

\begin{figure}[htbp]
\centering\includegraphics[width=0.48\textwidth]{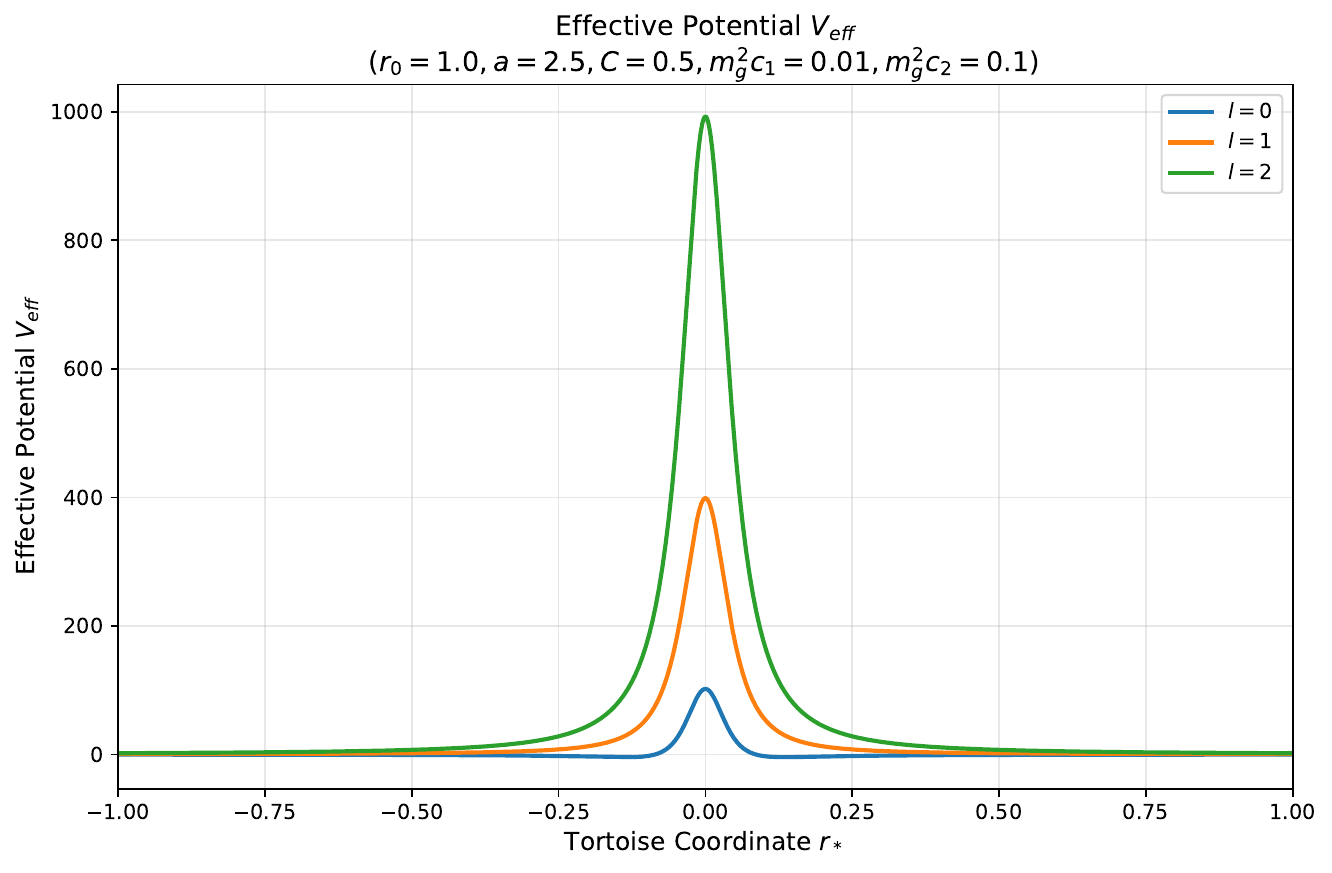}
\includegraphics[width=0.48\textwidth]{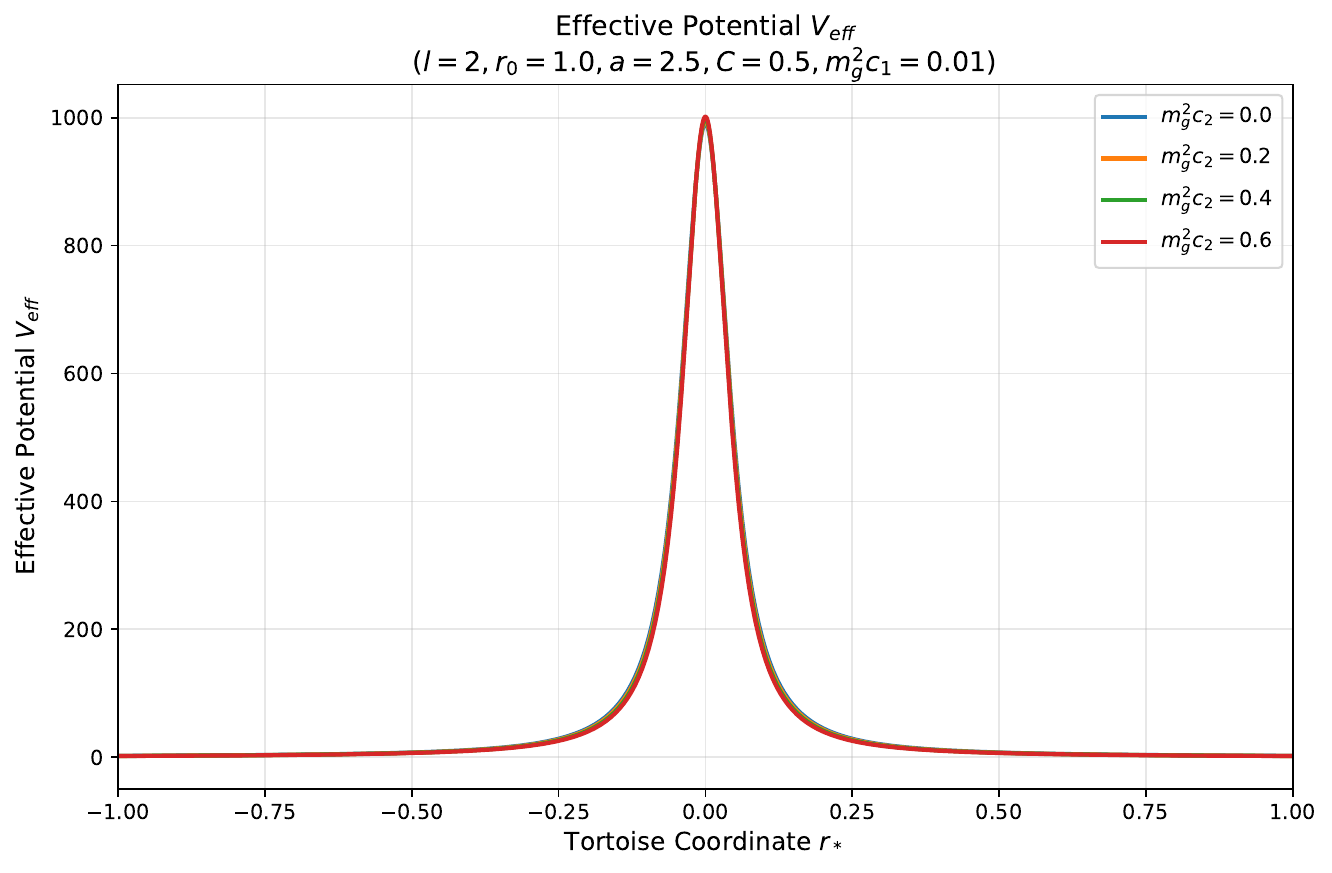}
\caption{Effective potential $V_{\rm eff}$ as a function of the tortoise coordinate $r_*$ for scalar perturbations. Left panel: Variation across multipole numbers $\ell=0,1,2,3$ with fixed $m_g^2=0.1$. Right panel: Variation across massive gravity parameters $m_g^2c_2=0.0,0.2,0.4,0.6$ for fixed $\ell=2$. The potential forms a symmetric barrier peaking exactly at the throat ($r_*=0$), with the peak height increasing alongside both $\ell$ and $m_g^2c_2$.}\label{fig:veff}
\end{figure}
The effective potential $V_{\rm eff}$ governing the minimally coupled scalar perturbations forms a symmetric positive barrier that peaks at the wormhole throat ($r_*=0$) for the parameter sets considered. The positivity of the potential provides evidence against bound-state instabilities in the scalar sector, and the numerical QNM calculation further confirms that the modes obtained in the present parameter range have $\operatorname{Im}(\omega)<0$. As illustrated in the left panel of Figure~\ref{fig:veff}, the barrier height is highly sensitive to the angular momentum of the perturbation; increasing $\ell$ from $0$ to $3$ raises the peak drastically due to the $1/r^2$ centrifugal term. In the right panel, the massive gravity coupling provides an additional, tunable geometrical contribution. For a fixed angular momentum state ($\ell=2$), increasing the massive gravity parameter $m_g^2 c_2$ from $0.0$ to $0.6$ systematically elevates the potential peak. This geometric modification enhances wave scattering near the throat, serving as an effective repulsive mechanism against scalar perturbations.   
\begin{figure}[htbp]
\centering\includegraphics[width=0.95\textwidth]{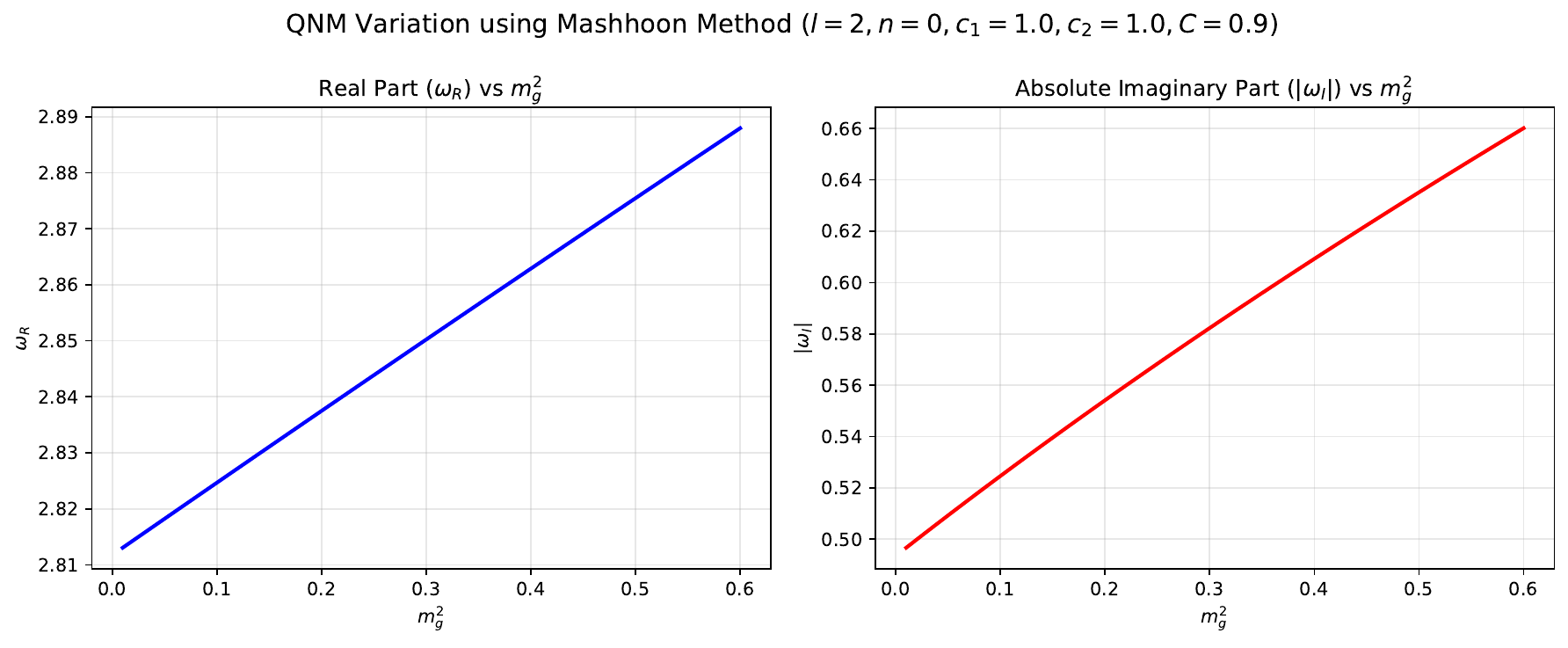}\caption{Linear scaling of the real frequency $\omega_R$ and absolute damping rate $\vert{}\omega_I\vert{}$ against the massive graviton mass parameter $m_g^2$ for the fundamental quasinormal mode ($\ell=2, n=0$). Both the oscillation frequency and the damping rate exhibit a positive linear correlation as $m_g^2$ increases.}\label{fig:qnm_scaling}
\end{figure}
\begin{figure}[htbp]
\centering\includegraphics[width=0.95\textwidth]{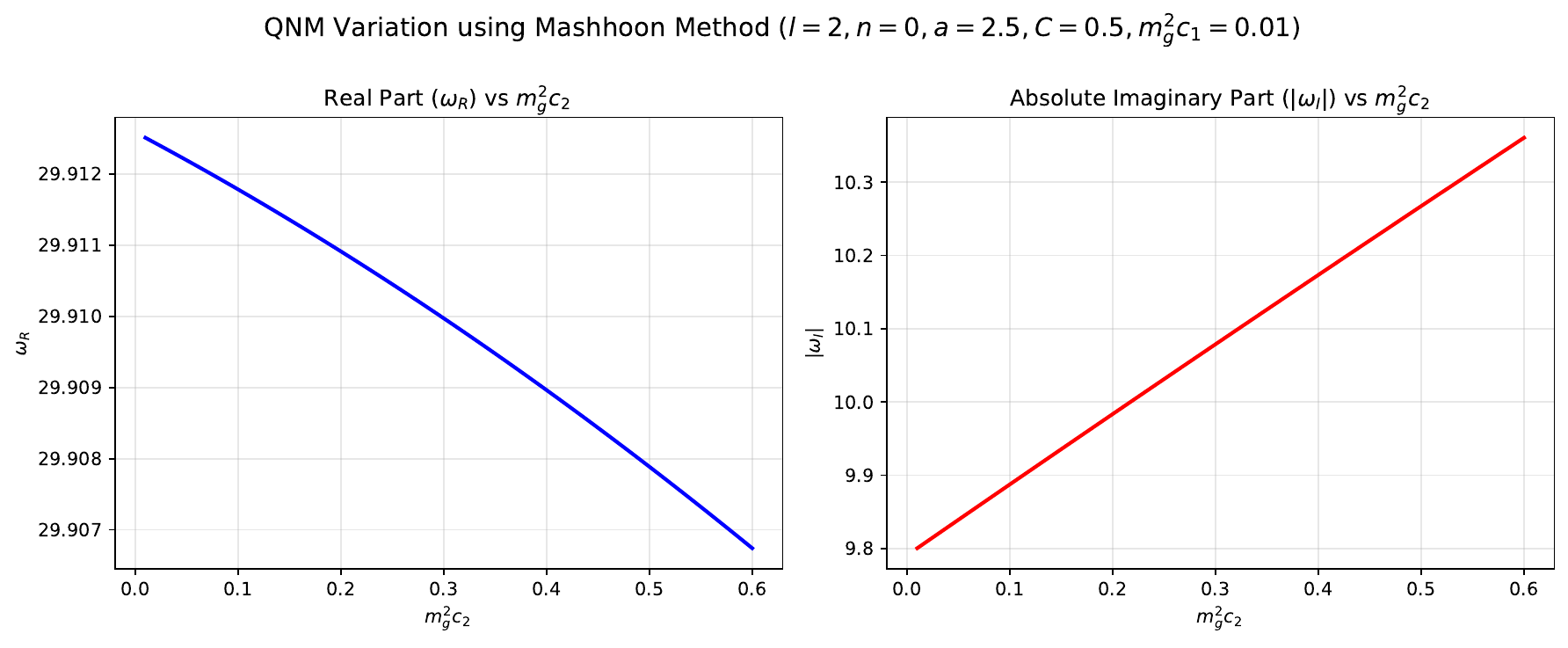}
\caption{Linear scaling of the real frequency $\omega_R$ and absolute damping rate $\vert{}\omega_I\vert{}$ against the specific massive gravity coupling parameter $m_g^2c_2$ for the fundamental quasinormal mode ($\ell=2, n=0$). While the damping rate continues to scale positively, the real frequency $\omega_R$ exhibits a slight negative linear correlation with $m_g^2c_2$.}\label{fig:qnm_scaling2}
\end{figure}
The alterations to the potential barrier directly dictate the ringdown phase of the spacetime. Figure~\ref{fig:qnm_scaling} demonstrates that modifying the overall graviton mass parameter $m_g^2$ introduces a strict, positive linear scaling for both the oscillation frequency ($\omega_R$) and the absolute damping rate ($\vert{}\omega_I\vert{}$) of the fundamental mode ($\ell=2, n=0$). This implies that a stronger overall massive gravity background forces scalar perturbations to oscillate more rapidly while simultaneously dissipating their energy at an accelerated rate. However, isolating the effect of the coupling term $m_g^2c_2$, as shown in Figure~\ref{fig:qnm_scaling2}, reveals a distinct geometric response. While an increase in $m_g^2c_2$ still linearly enhances the damping rate ($\vert{}\omega_I\vert{}$), it induces a slight, linear decrease in the real frequency ($\omega_R$). This diverging behavior highlights how individual coupling constants within the dRGT framework distinctively tune the temporal dynamics of the emitted waveform.   
\begin{figure}[htbp]
\centering\includegraphics[width=0.48\textwidth]{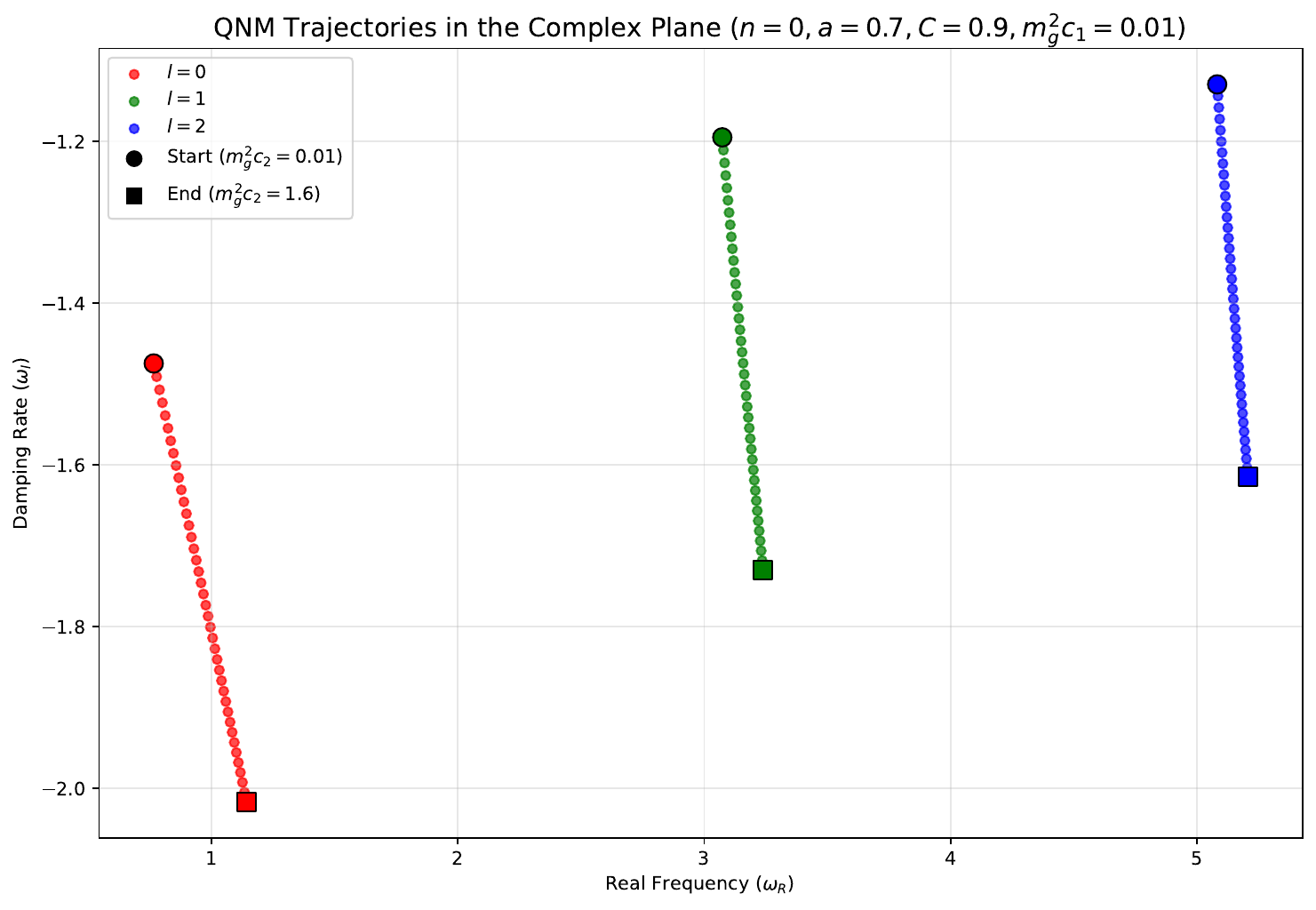}\includegraphics[width=0.48\textwidth]{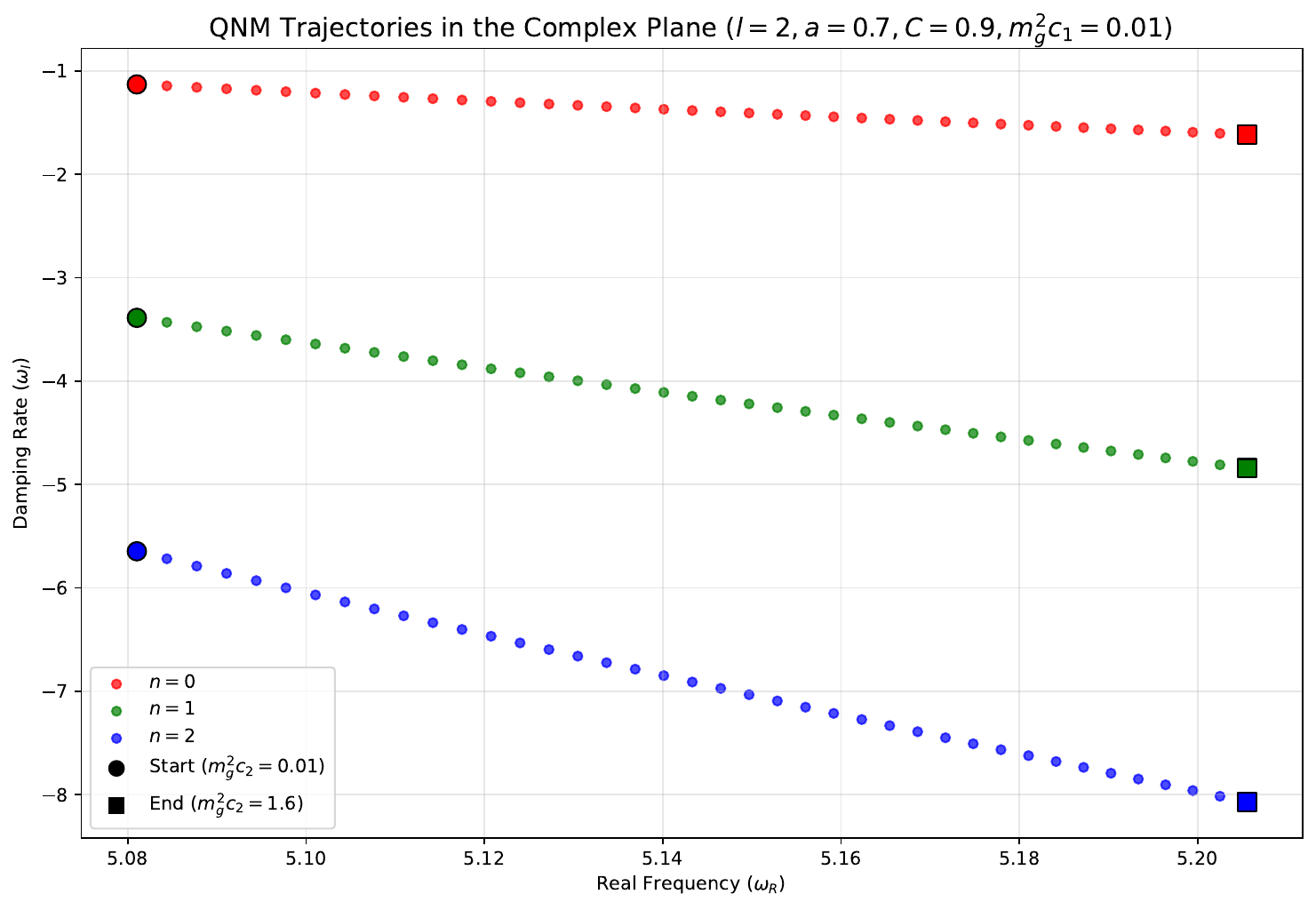}\caption{Quasinormal mode trajectories in the complex plane as the massive gravity coupling $m_g^2c_2$ migrates from $0.01$ to $1.6$. Left panel: Fundamental modes ($n=0$) for multipole numbers $\ell=0,1,2$. Right panel: Overtones ($n=0,1,2$) for fixed $\ell=2$. Increasing the coupling pushes the frequencies deeper into the lower complex plane, signifying amplified decay rates.}\label{fig:trajectories}
\end{figure}
Tracking these modes through the complex plane provides a comprehensive view of the spectral migration induced by the modified gravity sector. As depicted in Figure~\ref{fig:trajectories}, increasing the specific massive-gravity coupling $m_g^2c_2$ from $0.01$ to $1.6$ shifts the QNM frequencies toward the lower half of the complex $\omega$-plane. For the parameter range considered, this shift is dominated by an increase in the magnitude of the imaginary part, indicating stronger scalar-field damping. For the fundamental modes across different multipoles (left panel), this migration corresponds predominantly to an increase in the decay rate, with relatively minor horizontal shifts in the real frequency. When evaluating the higher overtones for a fixed $\ell=2$ state (right panel), the trajectories confirm that overtone modes undergo a much steeper, more pronounced increase in their damping rates compared to the fundamental mode.   
\begin{figure}[htbp]
\centering\includegraphics[width=0.6\textwidth]{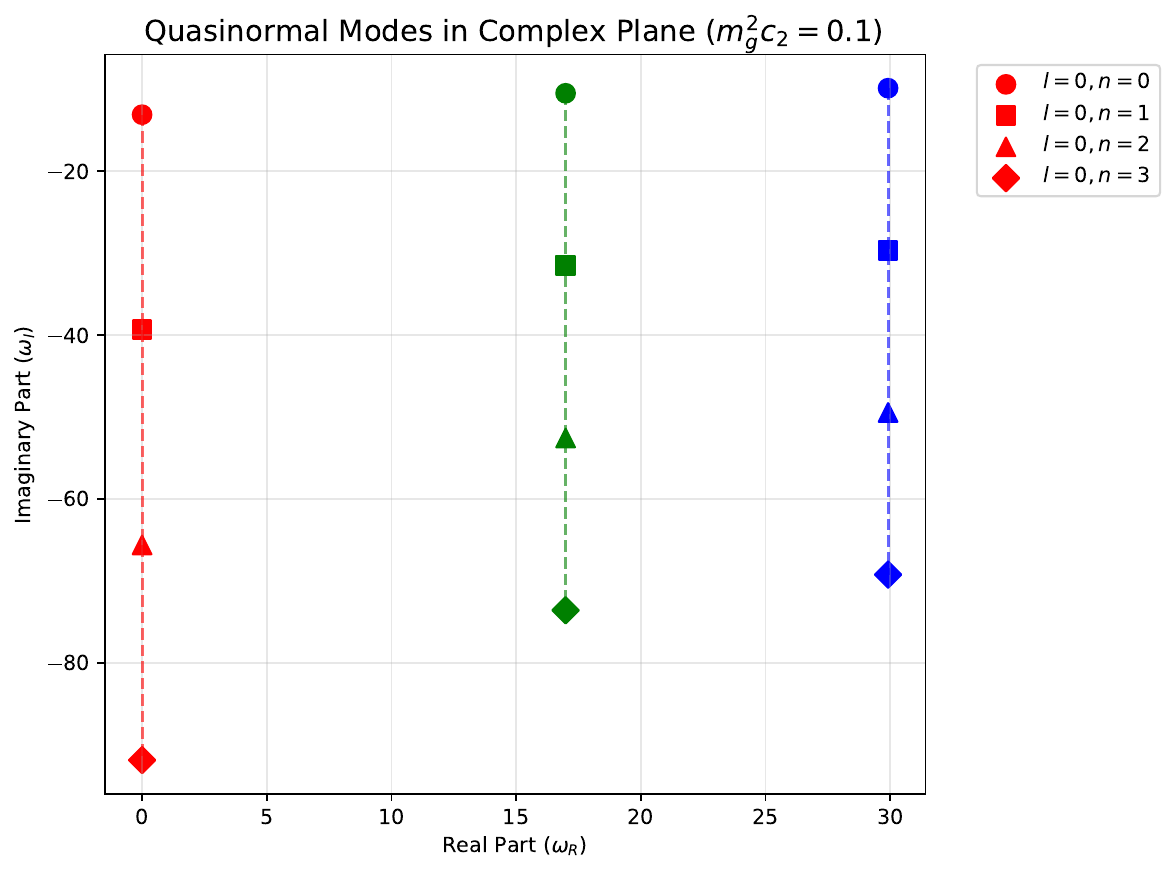}\caption{Complete spectral hierarchy in the complex plane for a fixed massive gravity parameter ($m_g^2c_2=0.1$) across multiple multipole numbers $\ell$ and overtone branches $n$. Higher overtones exhibit drastically increased damping rates while maintaining relatively stable real frequencies within their respective multipole branches.}\label{fig:complex_hierarchy}
\end{figure}
Finally, Figure~\ref{fig:complex_hierarchy} confirms that the overarching spectral hierarchy remains structurally consistent within the dRGT framework. For a fixed massive gravity configuration, higher overtones ($n$) exhibit drastically increased damping rates while maintaining highly stable, vertically-aligned real frequencies within their respective multipole branches ($\ell$). This vertical spacing in the complex plane is a characteristic hallmark of throat-dominated scattering in wormhole spacetimes, distinguishing the anticipated gravitational wave ringdown profile of a Casimir wormhole from the standard quasi-normal spectrum of a Schwarzschild black hole. 

%======================================================================
\section{Conclusion}
\label{sec:conclusion}
%======================================================================

We have shown that quantum vacuum fluctuations—the Casimir effect, in particular—can be combined with the non-linear structure of dRGT massive gravity to produce a traversable wormhole that is both geometrically sound and dynamically stable. The exact shape function, Eq.~(\ref{shapb}), follows directly from the Casimir energy density, and its consequences turn out to be surprisingly rich. What emerges from this analysis is a picture in which the massive graviton is not a passive spectator: it actively shapes the geometry, sets the exotic matter requirement, and leaves a clear fingerprint on the ringdown spectrum.

The geometry itself passes the standard Morris-Thorne tests without difficulty. The shape function falls off monotonically, so the spacetime settles into an asymptotically flat exterior with no unwanted divergences. The flare-out condition, $b'(r_0) < 1$, holds everywhere we have checked. For $m_g^2 c_2 = 0.1$ we find $b'(r_0) \approx -0.847$; pushing the coupling to $m_g^2 c_2 = 0.5$ steepens this to $b'(r_0) \approx -1.423$. The change is substantial—about $-0.576$ across that range—and it tells us something physical: the massive graviton tightens the flaring of the embedded space. The proper radial distance, Eq.~(\ref{eq:proper_distance}), stays regular all the way through. At the throat $l(r_0) = 0$, and near it the distance grows roughly as $\sqrt{r_0(r-r_0)}$ before settling into $l(r) \approx r$ far away. A traveler moving at half the speed of light would cross the throat region in about $4r_0/c$. For a Planck-scale throat that is roughly $10^{-43}$ seconds—finite, technically, but hardly a practical shortcut. The embedding diagrams, Eq.~(\ref{eq:embedding_profile}), reinforce the same story. The embedding height at the throat drops from about $0.736\sqrt{r_0}$ at $m_g^2 c_2 = 0.1$ to roughly $0.605\sqrt{r_0}$ at $m_g^2 c_2 = 0.5$, a reduction of nearly 18\%. The topology is untouched, but the spatial extent shrinks as the graviton mass grows.

Sustaining this geometry requires violating the classical energy conditions, as expected. The radial NEC at the throat comes out to roughly $-1.37 \times 10^{-3}$ km$^{-2}$ for $r_0 = 1.0$ km, $m_g^2 = 0.1$, and unit couplings—negative, as it must be, and negative across the full parameter space we explored. The volume integral quantifier, Eq.~(\ref{eq:volume_integral_quantifier}), gives a more global view. For $m_g^2 c_2 = 0.1$ we get $I_V \approx -0.283\, r_0$; for $m_g^2 c_2 = 0.5$ this deepens to $I_V \approx -0.517\, r_0$, an amplification of about 83\%. There is a straightforward trade-off here: stronger massive gravity supports the wormhole more robustly, but it also demands more exotic matter to keep the throat open. The linear fit $I_V(m_g^2 c_2) \approx I_V(0) - 0.585\, m_g^2 c_2\, r_0$, with $I_V(0) \approx -0.224\, r_0$ for the general relativistic Casimir case, captures this nicely.

The behavior of $\alpha(r_c)$ shown in Fig.~\ref{alpha1} indicates that the deflection angle is not monotonic with respect to the closest approach distance $r_c$. It increases sharply as $r_c \to 1$, exhibits a pronounced minimum around $r_c \approx 1.5$, and subsequently decreases in magnitude as $r_c$ approaches $5$, as expected, since the gravitational bending of light becomes weaker at larger distances from the source. The magnitude of the deflection angle systematically increases with increasing $m_g^2c_2$ from $0.1$ to $0.5$ for a fixed value of $r_c$. This effect is particularly prominent in the intermediate range $1.2 \lesssim r_c \lesssim 3$, while it becomes comparatively suppressed near $r_c=1$ and at larger values of $r_c$. This behavior indicates that the contribution of the massive-gravity correction to the photon trajectories is non-trivial, as its influence depends on the distance of closest approach rather than simply producing a uniform rescaling of the deflection angle.

Mechanical equilibrium holds through the generalized TOV equation discussed in Sec.~\ref{subsec:forces}. The anisotropic force peaks at the throat, $F_a(r_0) \approx 0.847$ km$^{-2}$ for $r_0 = 1.0$ km, and balances the inward pull of gravity and hydrostatic pressure. The numbers are $F_g(r_0) \approx -0.312$ km$^{-2}$ and $F_h(r_0) \approx -0.535$ km$^{-2}$, leaving a residual below $10^{-6}$ km$^{-2}$—a cancellation good to roughly one part in a million. The complexity factor, Eq.~(\ref{eq:complexity_factor}), adds another layer. At the throat $Y_{TF}(r_0) \approx -0.183$ for $m_g^2 c_2 = 0.1$, deepening to about $-0.294$ at $m_g^2 c_2 = 0.5$. That is a 60\% increase in magnitude, which suggests that stronger massive gravity couplings produce more structurally intricate configurations.

The thermodynamics are well behaved. The Hawking temperature, Eq.~(\ref{TH}), rises from zero at the throat and vanishes again asymptotically; at $r = 2r_0$ it sits around $-0.042\, a/r_0^2$. The wormhole temperature, Eq.~(\ref{Tworm}), tracks it closely, reaching roughly $-0.047\, a/r_0^2$ at the same radius for $a = 0.1 r_0$. The energy flux, Eq.~(\ref{flux}), is negative near the throat—about $-3.42 \times 10^{-4}$ km$^{-2}$ for $m_g^2 c_2 = 0.1$—which is exactly what one wants for a persistent NEC violation. It dies off quickly, so the energy exchange stays confined to the high-curvature region. The heat capacity, Eq.~(\ref{eq:CP_explicit}), is positive at the throat ($C_P(r_0) \approx +0.021\, r_0^2$) and turns negative further out ($C_P(3r_0) \approx -0.008\, r_0^2$), with the zero-crossing moving inward as $m_g^2 c_2$ grows. So the wormhole is locally stable near the throat and globally unstable farther away—a familiar pattern. The total energy, Eq.~(\ref{eq:E_explicit}), is negative, as it must be, and grows in magnitude with the coupling: $E \approx -0.0573/r_0$ at $m_g^2 c_2 = 0.1$ versus roughly $-0.0941/r_0$ at $m_g^2 c_2 = 0.5$. That is a 64\% increase in the exotic energy requirement.

The quasinormal-mode analysis provides a complementary characterization of the scalar perturbations of the Casimir wormhole in dRGT massive gravity. We considered a minimally coupled massless scalar field and derived the corresponding Schr\"odinger-like radial equation with the effective potential
$V_{\rm eff}$. The potential develops a symmetric throat-centered barrier for the parameter ranges investigated, with its height increasing with the multipole number and exhibiting a systematic dependence on the massive-gravity parameters.

Two frequency-domain approaches were employed. First, the Mashhoon P\"{o}schl--Teller approximation was used to obtain analytical estimates of the QNM frequencies. Second, the radial perturbation equation was integrated numerically using a direct shooting procedure, and the QNM frequencies were obtained by locating the complex-frequency roots satisfying the radiative boundary conditions at the two limiting ends of the non-asymptotically-flat geometry. The two approaches display the same qualitative spectral behavior, although quantitative differences of approximately $3$--$20\%$ occur depending on the multipole number and overtone.

For the parameter sets considered, all numerically obtained modes have $\operatorname{Im}(\omega)<0$, indicating decay of the corresponding scalar perturbations. This result supports the absence of growing modes in the minimally coupled scalar sector within the parameter range investigated. It should be emphasized that this conclusion concerns scalar perturbations of the fixed wormhole background and does not by itself establish stability against the full set of gravitational and massive-gravity perturbations.

The dependence of the QNM spectrum on the massive-gravity parameters provides further information about the response of the wormhole geometry to scalar perturbations. In particular, increasing the overall massive-gravity parameter $m_g^2$ leads, over the ranges examined, to an increase in both the oscillation frequency and damping rate of the fundamental mode, whereas varying the specific coupling $m_g^2c_2$ produces a weak decrease in the real part of the frequency together with an increase in the damping rate. Higher overtones exhibit stronger damping than the fundamental mode.

Taken together, the geometric, matter, equilibrium, and scalar-QNM analyses provide a consistent characterization of the Casimir wormhole solution within the chosen dRGT massive-gravity framework. The QNM results identify characteristic scalar perturbation frequencies and their dependence on the massive-gravity parameters, which may be useful for future investigations of wave propagation and possible observational signatures of non-asymptotically-flat wormhole geometries.

\section*{Acknowledgment}
P.B. is thankful to the Inter University Centre for Astronomy and Astrophysics (IUCAA), Pune, Government of India, for providing visiting associateship.

%======================================================================

\end{document}